\documentclass[aps,prl,twocolumn,floatfix,longbibliography,superscriptaddress]{revtex4-2}
\usepackage{bbm}
\usepackage{graphicx}
\usepackage{dcolumn}
\usepackage{bm}
\usepackage{subfigure}
\usepackage{amsmath}
\usepackage{feynmf}
\usepackage{hyperref}
\usepackage{amssymb}
\usepackage{attachfile}
\usepackage{multirow}
\usepackage{makecell}
\newcommand{\bk}{\boldsymbol k}

\newcommand{\bq}{\boldsymbol q}

\newcommand{\zb}{\color {black}}

\usepackage{braket}
\usepackage{esint}
\usepackage{times}

\begin{document}

\title{Stripe-Order Altermagnetism: Nematic Spin Splitting beyond the \texorpdfstring{$l$}{l}-Wave Classification}

\author{Zheng-Yang Zhuang}
\affiliation{Guangdong Provincial Key Laboratory of Magnetoelectric Physics and Devices,
	State Key Laboratory of Optoelectronic Materials and Technologies,
	School of Physics, Sun Yat-sen University, Guangzhou 510275, China}

\author{Zhongbo Yan}
\email{yanzhb5@mail.sysu.edu.cn}
\affiliation{Guangdong Provincial Key Laboratory of Magnetoelectric Physics and Devices,
	State Key Laboratory of Optoelectronic Materials and Technologies,
	School of Physics, Sun Yat-sen University, Guangzhou 510275, China}

\date{\today}

\begin{abstract}
Altermagnetism combines compensated magnetic order with nonrelativistic spin splitting, yet established mechanisms predominantly rely on spin-reversing rotations in Néel-order antiferromagnets. Here we establish stripe-order altermagnetism governed instead by a spin-reversing mirror, placing it outside the usual rotation-based $l$-wave classification.
Using two-orbital models, we show that the interplay between stripe spin and orbital orders can yield a stripe altermagnet with mirror-constrained nematic spin splitting or a stripe anti-altermagnet with spin-degenerate bands. The latter can support ferroelectric-like electrical control of spin splitting in suitable buckled structures. Random-phase-approximation (RPA) calculations show that stripe-order altermagnetism is favored near half filling under strong hopping anisotropy. The spin-reversing mirror further enforces a purely transverse Drude spin current for an electric field parallel or normal to the mirror plane, while spin-resolved mirror and rotation probes distinguish this response from that of rotation-governed $l$-wave altermagnets.
\end{abstract}

\maketitle

\begin{figure}[t]
	\includegraphics[width=\columnwidth]{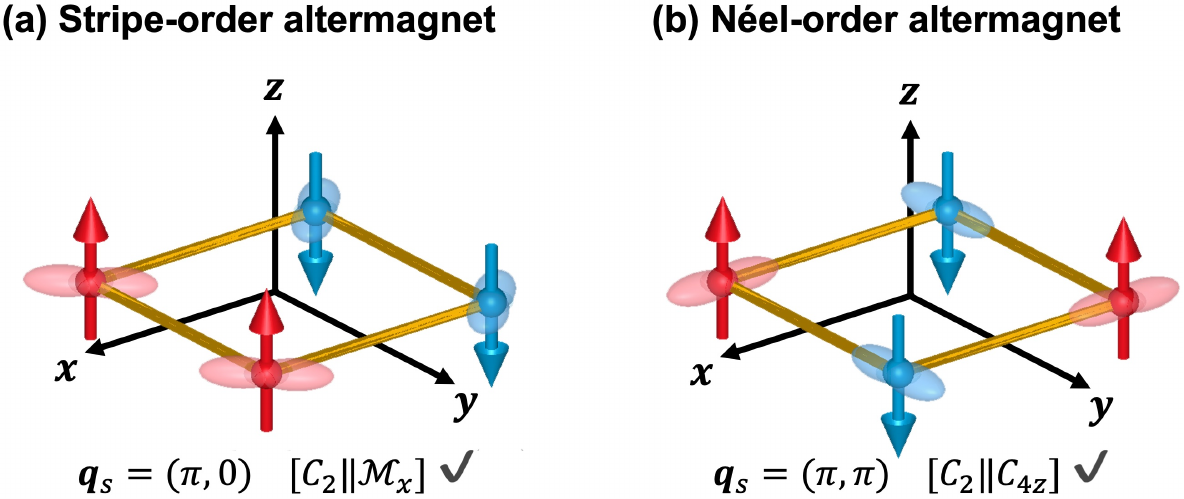}
		\caption{Real-space comparison of (a) stripe-order and (b) Néel-order altermagnets. Red and blue arrows denote opposite local moments, and translucent lobes represent anisotropic orbital densities. The stripe and Néel configurations have ordering vectors $\bq_s=(\pi,0)$ and $\bq_s=(\pi,\pi)$, respectively. For the spin patterns alone, the Néel geometry admits spin-reversing mirror $[C_{2}\Vert\mathcal{M}_{i}]$ and rotational $[C_{2}\Vert C_n]$ relations, whereas the ferromagnetic-chain geometry of stripe order naturally admits a mirror relation between chains carrying opposite moments. Orbital order selects which relations remain symmetries of the full pattern.}
	\label{fig1}
\end{figure}

Unconventional magnets can combine antiferromagnetic spin order with responses conventionally associated with ferromagnets~\cite{Hayami2019NRSS,Hayami2020NRSS,Liu2025}. Altermagnetism (AM) is a prominent example~\cite{Ma2021AM,Libor2022AMa}: it supports spin-split bands in a collinear magnet with vanishing net spin magnetization~\cite{Yuan2020AM,Libor2022AMa,Yuan2021AM}, thereby reconciling antiferromagnetic compensation with momentum-dependent spin splitting.
Such nonrelativistic spin splitting was anticipated in $d$-wave spin-density-wave states~\cite{Hirsch1990,Ikeda1998} and spin-channel Pomeranchuk instabilities~\cite{wu2004,wu2007}, and was later recognized as a generic feature of a broad class of antiferromagnets in which a spin-reversing rotation or mirror, rather than inversion or translation, relates the opposite-moment sublattices~\cite{Ma2021AM,Libor2022AMa}. It has since been observed in CrSb~\cite{Reimers2024,Ding2024CrSb,Yang2024CrSb,Zeng2024CrSb,Li2024CrSb,Lu2024AM7}, MnTe~\cite{Osumi2024MnTe,Lee2024MnTe,Krempasky2024,Hajlaoui2024AM}, and KV$_2$Se$_2$O~\cite{Jiang2024KV2Se2O,Zhang2025Cpair,wang2025KVSeO,Jiao2026KVSeO}. Altermagnetism has also been linked to diverse phenomena across condensed matter physics~\cite{Libor2022AMb,chen2026review}, most notably spintronics~\cite{Bai2023AM,Han2024AM,Hu2025NLME,Duan2025AFMAM,Gu2025FEAM,Jungwirth2025review}, topological phases~\cite{Zhu2023TSC,Zhu2024dislocation,Li2023AMHOTSC,Li2024AMHOTI,Ghorashi2024AM,Antonenko2024AM}, and unconventional superconductivity~\cite{Zhang2024AM,Brekke2023AM,Monkman2026,Fukaya2025review,Liu2025review}. 
Altermagnetic spin splittings are commonly classified by the angular-harmonic degree $l$ as $p$-wave ($l=1$)~\cite{Hellenes2023pwave,Brekke2024pwave,Yu2025Odd,liu2025floquet,huang2025oddparityAM}, $d$-wave ($l=2$)~\cite{Libor2022AMa}, $f$-wave ($l=3$)~\cite{lin2025,zeng2025,huang2025oddparityAM,zhu2025floquet,li2025floquet,zhuang2025,Pan2025odd}, or higher-$l$ forms. These labels are assigned according to how the spin splitting transforms under spin-reversing rotations, which combine a real-space rotation with spin reversal and thereby constrain its angular dependence.
This framework is prevalent in part because many widely studied altermagnets are Néel-order antiferromagnets (Néel AFMs), whose spin structures naturally accommodate such rotations [Fig.~\ref{fig1}(b)]. In spin-space-group notation~\cite{Liu2022AM,Libor2022AMa,Liu2024AMPRX,Jiang2024SSG,Xiao2024SSG,liu2026OSSG}, these symmetries are denoted $[C_2\Vert C_n]$, with $C_2$ reversing spin and $C_n$ an $n$-fold real-space rotation. Although spin-reversing mirrors have also been recognized as supporting altermagnetism, their independent role in characterizing spin splitting remains underexplored because they typically coexist with spin-reversing rotations in Néel-order altermagnets (Néel AMs).

Alongside Néel order, stripe order constitutes another important class of collinear magnetic order widely studied in correlated materials~\cite{Zaanen1989stripe,Machida1989,Kato1990,hucker2012stripe,tranquada1995stripe,Kivelson2003stripe,Missiaen2025stripe,Chen2008nematic,Xu2008stripe,Fernandes2012nematic,Achkar2016stripe}. Unlike Néel order, in which local moments alternate along every nearest-neighbor direction, stripe order consists of one-dimensional ferromagnetic chains stacked antiferromagnetically [Fig.~\ref{fig1}(a)]. Among crystallographic spin-reversing rotations $[C_{2}\Vert C_{nz}]$, the two-dimensional stripe geometry admits at most $[C_2\Vert C_{2z}]$. Its presence or absence determines whether the $l=1$ rotational label is available, so the stripe case reduces to a binary ($\mathbb Z_2$) distinction. When spinless time-reversal symmetry $[\bar C_2\Vert\mathcal T]$ is also preserved, however, $[C_2\Vert C_{2z}]$ enforces spin degeneracy (see End Matter). Therefore, when altermagnetism appears in stripe AFMs that preserve $[\bar C_2\Vert\mathcal T]$, the resulting phases---hereafter termed \textit{stripe-order altermagnets} (stripe AMs)---must lack every such spin-reversing rotation and therefore lie outside the usual rotation-based angular-harmonic classification.
By contrast, stripe order can admit a spin-reversing mirror $[C_{2}\Vert\mathcal{M}_{i}]$ that exchanges chains with opposite moments and constrains the spin splitting. This motivates a distinct mirror-governed class of altermagnetism.

In this Letter, we introduce altermagnetism into stripe-order AFMs and establish stripe-order altermagnetism as a realization of mirror-governed \textit{nematic altermagnetism}, a class beyond the rotation-based $l$-wave framework. We use this term for altermagnetic states in which a spin-reversing mirror, but no spin-reversing rotation, relates the opposite-spin sectors and constrains their momentum-space spin splitting. Here ``nematic'' labels the symmetry of the spin-splitting texture rather than an additional order parameter. Symmetry analysis shows that orbital order provides a route to stripe AMs and electrically controllable \textit{stripe anti-altermagnets} (stripe anti-AMs). RPA calculations identify the conditions favorable to stripe AM, while spin-current responses provide transport probes that distinguish mirror-governed nematic altermagnets from rotation-governed $l$-wave altermagnets.

{\it \color{blue}Symmetry and models for stripe AMs.--} An altermagnet features spin-split bands yet zero net spin magnetization~\cite{Libor2022AMa,Ma2021AM}. This requires two conditions. First, the two magnetic sublattices with opposite moments must experience inequivalent local environments; hence spin-reversing inversion $[C_{2}\Vert\mathcal{P}]$ and translation $[C_{2}\Vert t]$ symmetries must be absent. Second, the opposite-moment sublattices must still be related by a spin-reversing symmetry.
In a stripe AFM, this relation can be provided by a vertical mirror such as $\mathcal{M}_{x}$ or $\mathcal{M}_{y}$.
Within the two-dimensional stripe geometry considered here, a minimal criterion for realizing a stripe AM is the absence of $[C_{2}\Vert\mathcal{P}]$ (equivalently $[C_{2}\Vert C_{2z}]$) and $[C_{2}\Vert t]$, together with a preserved spin-reversing mirror.

\begin{figure}[t]
		\includegraphics[width=\columnwidth]{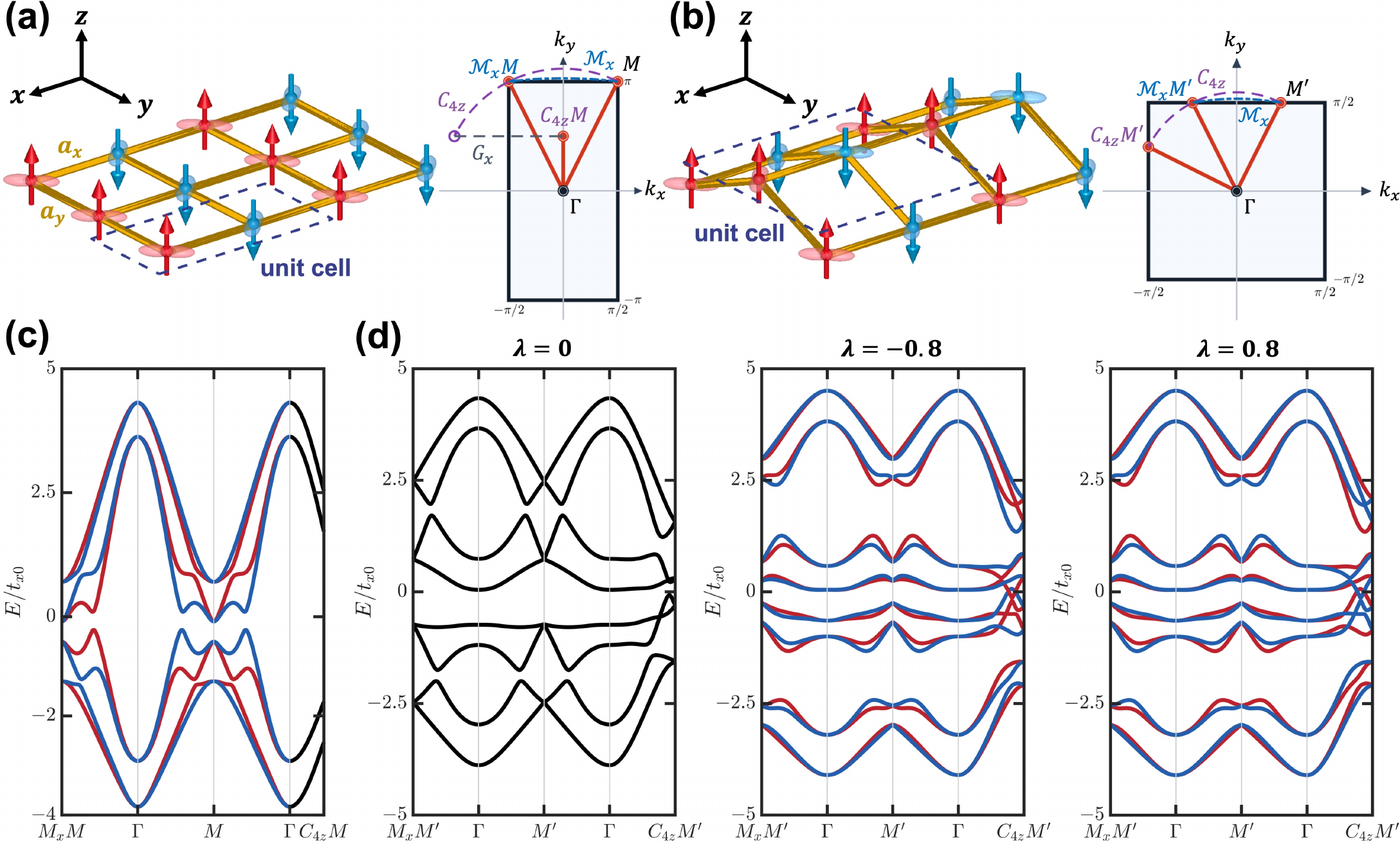}
			\caption{Stripe-altermagnetic models and spin-resolved band structures. (a) Minimal orbital-order stripe-AM model and its reduced Brillouin zone. (b) Buckled stripe-spin/Néel-$\tau_x$-orbital model and its reduced Brillouin zone. The labels $\mathcal{M}_xM$ and $C_{4z}M$ denote the images of $M$ under the corresponding operations, and $G_x=\pi$ is the reciprocal-lattice period along $x$. Panels (c) and (d) show spin-resolved bands along the paths in (a) and (b), respectively. In (d), the spectra from left to right correspond to $\lambda=0,-0.8,+0.8$. Finite $\lambda$ activates the spin splitting, and $\lambda\rightarrow-\lambda$ reverses its pattern. Red and blue denote opposite spins, whereas black denotes spin-degenerate bands. The opposite-spin bands are related by the spin-reversing mirror $[C_{2}\Vert\mathcal{M}_{x}]$, rather than by the spin-reversing rotation $[C_{2}\Vert C_{4z}]$. We set the orbital exchange fields $\delta_x$ and $\delta_g$ to $0.6$ in the stripe-AM and stripe-spin/Néel-orbital models, respectively. The remaining parameters are $(t_{x0},t_{x1},t_{y0},t_{y1},t_{d0},t_{d1},J)=(1,0.2,0.15,0.03,0.4,0.3,0.4)$.}
		\label{fig2}
	\end{figure}

Because stripe order often coexists with orbital order~\cite{Fernandes2012nematic}, we use the latter to break spin-reversing inversion $[C_{2}\Vert\mathcal{P}]$ and translation $[C_{2}\Vert t]$. Orbital-order-induced altermagnetism has also been explored in Néel AFMs~\cite{Leeb2024AM,Vila2025orbitalorder,Meier2026,zhuang2025}. In momentum space,
the Hamiltonian takes the general form
\begin{equation}
	\mathcal{H}(\bk)=\mathcal{H}_{\rm ele}(\bk)s_{0}
	+\mathcal{H}_{J}s_{z}+\mathcal{H}_{\rm orb}s_{0}.
\end{equation}
Here, $s_0$ is the identity matrix and $\boldsymbol{s}=(s_{x},s_{y},s_{z})$ denotes the Pauli matrices in spin space. The terms $\mathcal{H}_{\rm ele}(\bk)$, $\mathcal{H}_{J}$, and $\mathcal{H}_{\rm orb}$ describe spin-independent hopping, stripe spin order, and orbital order, respectively. For the orbital-order mechanism considered here, a minimal model contains two sublattices, two orbitals, and two spin sectors, giving an eight-band Hamiltonian. For concreteness, we consider a two-dimensional square-lattice model with $d_{xz}$ and $d_{yz}$ orbitals [Fig.~\ref{fig2}(a)],
\begin{equation}
	\mathcal{H}_{\rm ele}(\bk)=
	\begin{pmatrix}
		\Delta_y(\bk) & \Delta_x(\bk)+\Delta_d(\bk)\\
		\Delta_x^{\dagger}(\bk)+\Delta_d^{\dagger}(\bk) & \Delta_y(\bk)
	\end{pmatrix}.
	\label{eq: stripeAM ele}
\end{equation}
The Pauli matrices $\boldsymbol{\tau}$ and $\boldsymbol{\sigma}$ act in orbital and sublattice spaces, respectively. We define $\Delta_x(\bk)=2\cos k_x(t_{x0}\tau_0+t_{x1}\tau_z)$, $\Delta_y(\bk)=2\cos k_y(t_{y0}\tau_0-t_{y1}\tau_z)$, and $\Delta_d(\bk)=4t_{d0}\cos k_x\cos k_y\tau_0-4t_{d1}\sin k_x\sin k_y\tau_x$ (the lattice constants $a_{x}$ and $a_{y}$ are set to unity throughout). These three terms describe, respectively, nearest-neighbor hopping along $x$, nearest-neighbor hopping along $y$, and next-nearest-neighbor hopping, with corresponding amplitudes $\{t_{x0},t_{x1}\}$, $\{t_{y0},t_{y1}\}$, and $\{t_{d0},t_{d1}\}$. The stripe spin order is $\mathcal{H}_{J}=J\tau_0\sigma_z$. Stripe orbital order with the same ordering vector is described by $\mathcal{H}_{\rm orb}=(\delta_x\tau_x+\delta_z\tau_z)\sigma_z$, where the terms $\delta_x\tau_x$ and $\delta_z\tau_z$ describe $d_{xz}\pm d_{yz}$ orbital antibonding/bonding and $d_{xz}/d_{yz}$ orbital polarization, respectively.

Before orbital order develops ($\delta_x=\delta_z=0$), the system preserves both $[C_{2}\Vert\mathcal{P}]$ and $[C_{2}\Vert t]$. Together with spinless time reversal, the former enforces spin degeneracy, while the latter does so directly. Orbital order removes these constraints, with the resulting magnetic state determined by the orbital channel. At finite $\delta_x$ and $\delta_z=0$, the $d_{xz}\pm d_{yz}$ order preserves $[C_{2}\Vert\mathcal{M}_{x}]$, which relates the two spin sectors and enforces vanishing net spin magnetization, while no remaining symmetry forces spin degeneracy. The system therefore realizes a stripe AM with mirror-constrained nematic spin splitting~\cite{Camerano2025}. By contrast, finite $\delta_z$ at $\delta_x=0$ breaks $[C_{2}\Vert\mathcal{M}_{x}]$. The spin sectors are then no longer related by symmetry, and a metallic system generally develops finite spin magnetization, corresponding to a weak ferrimagnetic state.

Figure~\ref{fig2}(c) illustrates how mirror-constrained spin splitting in stripe AMs differs from its rotationally protected counterpart: opposite-spin bands are related by a mirror rather than a spin-reversing rotation. The resulting nematic spin splitting lies outside the rotation-based $l$-wave classification because no spin-reversing rotation assigns it a global $l$-wave label. This orbital-order realization also motivates a minimal four-band model, presented in End Matter.

{\it \color{blue}Stripe-order anti-altermagnets.--} The construction above corresponds to the first of three orbital configurations relative to stripe spin order: (i) stripe orbital order with the same ordering vector as the spin order, (ii) stripe orbital order with an ordering vector orthogonal to that of the spin order, and (iii) Néel orbital order. For $\tau_x$ orbital-bonding order, cases (ii) and (iii) instead realize stripe-order anti-altermagnetic states~\cite{Meier2026} (also referred to as hidden-altermagnetic states~\cite{Matsuda2025}), whose hidden altermagnetic sectors can be revealed by an electric field. We focus below on case (iii); case (ii) is discussed in the Supplemental Material~\cite{supplemental}.

Combining Néel orbital order with stripe spin order enlarges the unit cell from two to four sublattices [Fig.~\ref{fig2}(b)], giving a sixteen-band model after including spin. We denote this stripe (C-type)-spin/Néel (G-type)-orbital configuration by the superscript ${\rm CG}$. The spin-independent hopping is $\mathcal{H}_{\rm ele}^{\rm CG}(\bk)=2\cos k_x(t_{x0}+t_{x1}\tau_z)\sigma_x+2\cos k_y(t_{y0}-t_{y1}\tau_z)\rho_x+4(t_{d0}\cos k_x\cos k_y-t_{d1}\sin k_x\sin k_y\tau_x)\rho_x\sigma_x$, whereas the stripe spin and Néel orbital orders are $\mathcal{H}_{J}=J\sigma_z$ and $\mathcal{H}_{\rm orb}=\delta_g\rho_z\sigma_z\tau_x$, respectively. The Pauli matrices $\boldsymbol{\sigma}$ and $\boldsymbol{\rho}$ act on the two sublattice degrees of freedom along $x$ and $y$, respectively. As illustrated in Fig.~\ref{fig2}(b), this model can be viewed in real space as two copies of the minimal stripe-AM model in Fig.~\ref{fig2}(a), with opposite orbital-order patterns, $d_{xz}\pm d_{yz}$ and $d_{xz}\mp d_{yz}$, connected by a half-unit-cell translation along $y$. An electrostatic potential imbalance between the two copies is described by $\lambda\rho_z$. Such an imbalance can arise from stripe charge order with an ordering vector orthogonal to the spin order. Alternatively, in the buckled geometry of Fig.~\ref{fig2}(b), ferroelectric polarization can induce this imbalance, which can then be tuned by a vertical field with $\lambda=eE_zd/2$ without breaking $[C_2\Vert\mathcal{M}_x]$. Here $e$ is the elementary charge and $d$ is the vertical separation between the two copies.

At $\lambda=0$, spin-reversing inversion $[C_{2}\Vert\mathcal{P}]$ relates the two copies and, together with spinless time reversal, enforces same-momentum spin degeneracy. The full band structure is therefore spin degenerate, even though each constituent copy is stripe altermagnetic. We refer to this state as a stripe anti-{\zb AM}. Finite $\lambda$ breaks $[C_{2}\Vert\mathcal{P}]$ and shifts the energies of the two copies in opposite directions, producing momentum-dependent spin splitting. The state thereby becomes altermagnetic because $[C_{2}\Vert\mathcal{M}_{x}]$ continues to enforce vanishing net spin magnetization. In the buckled geometry of Fig.~\ref{fig2}(b), a vertical electric field controls $\lambda$. Reversing the field sends $\lambda\rightarrow-\lambda$ and reverses the spin-splitting pattern [Fig.~\ref{fig2}(d)] without reversing the underlying spin or orbital order. The stripe-spin/Néel-orbital system thus realizes electrically switchable altermagnetism: the spin splitting vanishes at $E_z=0$, emerges at finite $E_z$, and reverses with the field direction.

Having established these mechanisms, we next use RPA to identify the conditions favorable to stripe AM.

\begin{figure}[t]
		\includegraphics[width=\columnwidth]{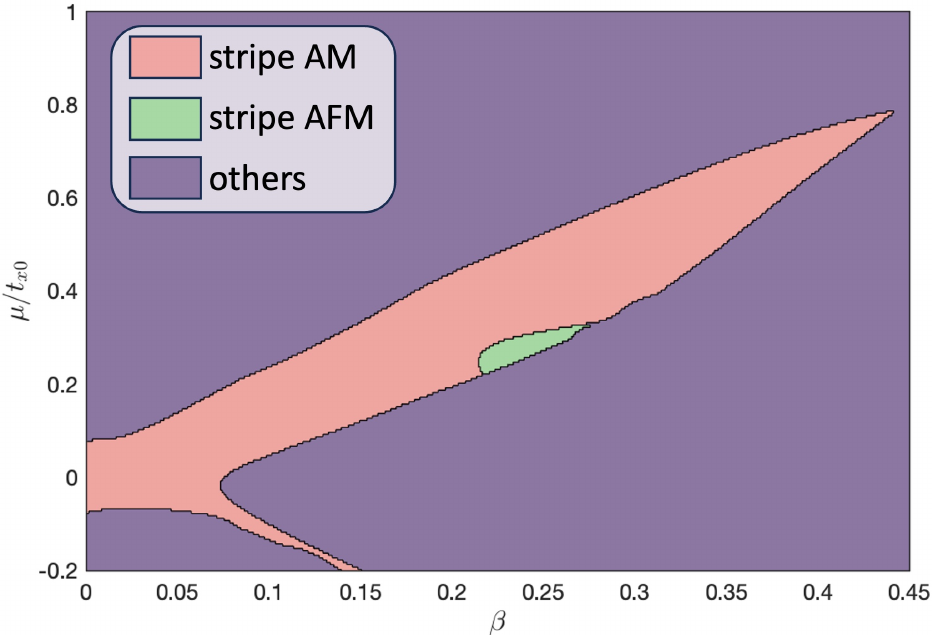}
		\caption{RPA leading-instability diagram as a function of the chemical potential $\mu/t_{x0}$ and hopping anisotropy $\beta\equiv t_{y0}/t_{x0}=t_{y1}/t_{x1}$. The effective interaction strengths in the spin and orbital channels are fixed at $U_s/t_{x0}=2.7$ and $U_o/t_{x0}=2.8$, respectively, and $k_BT/t_{x0}=0.04$. The label ``stripe AM'' denotes points at which the leading spin and orbital channels match the stripe-AM configuration, whereas ``stripe AFM'' denotes a stripe AFM with spin-degenerate bands that does not belong to the anti-AM class. The label ``others'' collects parameter points for which at least one active channel has its global maximum outside the set $\{\Gamma_q,X_q,Y_q,M_q\}$. Parameters: $\{t_{x0},t_{x1},t_{d0},t_{d1}\}=\{1,0.2,0.4,0.3\}$.}
	\label{fig3}
\end{figure}

{\it \color{blue}Spontaneous stripe AM.--} Motivated by the occurrence of stripe order in correlated materials, we use RPA to determine the conditions under which interactions favor the spin and orbital orders required for stripe AM.

Before either order develops, the parent model has no enlarged unit cell and can be written as a single-sublattice two-orbital model (see Supplemental Material~\cite{supplemental}). The ordering vector $\bq_{s/o}$ of the incipient spin or orbital order is identified by the momentum at which the corresponding susceptibility first diverges.
We define $\Gamma_q=(0,0)$, $X_q=(\pi,0)$, $Y_q=(0,\pi)$, and $M_q=(\pi,\pi)$, corresponding to ferromagnetic, stripe, stripe, and Néel order, respectively. Stripe spin order requires $\bq_s\in\{X_q,Y_q\}$. Within the present setting, stripe AM requires $\bq_o=\bq_s$ [case (i)], whereas stripe anti-AM requires either $\bq_o\in\{X_q,Y_q\}$ with $\bq_o\neq\bq_s$ [case (ii)] or $\bq_o=M_q$ [case (iii)]. Both phases considered here require $\tau_x$ orbital-bonding order.

We introduce effective onsite interactions in the spin and orbital channels and define normalized orbital matrices $\bar{\tau}_a=\tau_a/\sqrt{2}$. The projected susceptibilities are evaluated on a discrete mesh $\mathcal{D}_q$ covering $0\leq q_x,q_y\leq\pi$. After tracing over spin, the static bare susceptibility is
\begin{align}
		\chi^0_{\nu}(\bq)&=
		2\sum_{\bk,m,n}w_{\bk}
	\frac{f[E_n(\bk+\bq)]-f[E_m(\bk)]}
	{E_m(\bk)-E_n(\bk+\bq)}
	\nonumber\\
	&\quad\times
	\bigl|\langle u_m(\bk)\vert \mathcal{O}_\nu
	\vert u_n(\bk+\bq)\rangle\bigr|^2,
	\label{eq:lindhard}
\end{align}
where $\nu=s,(o,x),(o,z)$ labels the spin, orbital-bonding ($\tau_x$), and orbital-polarization ($\tau_z$) channels, with $\mathcal{O}_{\nu}=\bar{\tau}_0,\bar{\tau}_x,\bar{\tau}_z$, respectively; $w_{\bk}=1/(N_{k_x}N_{k_y})$; and $f(E)=[e^{(E-\mu)/(k_BT)}+1]^{-1}$ is the Fermi-Dirac distribution. The factor of two accounts for spin degeneracy. For each channel, the RPA-renormalized susceptibility is
$
	\chi^{\mathrm{RPA}}_{\nu}(\bq)
	=\frac{\chi^0_{\nu}(\bq)}{1-U_{\nu}\chi^0_{\nu}(\bq)}
$, while
$
	U_{c,\nu}(\bq)=\frac{1}{\chi^0_{\nu}(\bq)}$ is the critical interaction strength.
Here $U_\nu=U_s$ for spin and $U_\nu=U_o$ for the two orbital channels. We identify the leading instability from the global maximum
$\chi_\nu^{0,*}\equiv\max_{\bq\in\mathcal{D}_q}\chi_\nu^0(\bq)$
and apply the Stoner criterion $U_\nu\chi_\nu^{0,*}\geq1$.

As shown in Fig.~\ref{fig3}, the leading spin and orbital Stoner instabilities select the ingredients of stripe AM near half filling in the strongly anisotropic regime. Heuristically, the dominant $x$-direction hopping produces a quasi-one-dimensional dispersion and enhances nesting near $q_x=\pi$, favoring antiferromagnetic modulation along $x$, consistent with the $X_q=(\pi,0)$ stripe instability found numerically. As the hopping anisotropy is reduced, the stripe-AM region extends into the electron-doped regime. When only the effective onsite interactions are included, RPA preserves the locations of the bare-susceptibility maxima, which generally coincide in the spin and orbital channels for the present bands. Neither stripe anti-AM configuration is therefore selected. Nonlocal, orbital-dependent, or lattice-mediated interactions can modify this selection and may favor stripe anti-AM configurations in other regimes. The broad region labeled \emph{others} reflects cases in which at least one leading instability occurs at a generally incommensurate wave vector outside $\{\Gamma_q,X_q,Y_q,M_q\}$, as is common in quasi-one-dimensional bands.

{\it \color{blue}Spin-current response.--} Nematicity controls not only the spin-splitting pattern of stripe AMs but also their transport response. In the absence of spin-orbit coupling, the preserved spinless time-reversal symmetry $[\bar{C}_{2}\Vert\mathcal{T}]$ forbids the intrinsic linear charge Hall response, leaving the dissipative Drude term as the leading dc contribution. For a fixed spin sector $s$, the Drude charge-conductivity tensor is~\cite{Xiao2010review,Yan2017PRLMn3Ir}
\begin{align}
	\sigma^{c,s}_{ij}(\mu)
	=e^2\tau\sum_n\int_{\mathrm{RBZ}}\frac{d^2k}{(2\pi)^2}
	\left(-\frac{\partial f}{\partial E}\right)
	v_{i,n s}(\bk)v_{j,n s}(\bk),
	\label{eq:spin_resolved_drude}
\end{align}
where $\tau$ is the relaxation time and $v_{i,ns}(\bk)=\hbar^{-1}\partial_{k_{i}}E_{ns}(\bk)$ is the group velocity along the $i$th direction for the $n$th band in spin sector $s$.

The spin-reversing mirror constrains the two spin-resolved charge tensors according to
\begin{align}
	\sigma^{c,\downarrow}=R_x\sigma^{c,\uparrow}R_x^T,\qquad
	R_x=\mathrm{diag}(-1,1).
	\label{eq:mirror_tensor_constraint}
\end{align}
Accordingly, the two tensors take the form
\begin{align}
	\sigma^{c,\uparrow}=
	\begin{pmatrix}\sigma_{xx}^{c,\uparrow}&\sigma_{xy}^{c,\uparrow}\\ \sigma_{xy}^{c,\uparrow}&\sigma_{yy}^{c,\uparrow} \end{pmatrix},
	\quad
	\sigma^{c,\downarrow}=
	\begin{pmatrix}\sigma_{xx}^{c,\uparrow}&-\sigma_{xy}^{c,\uparrow}\\ -\sigma_{xy}^{c,\uparrow}&\sigma_{yy}^{c,\uparrow} \end{pmatrix}.
	\label{eq:spin_resolved_tensor_form}
\end{align}
The total charge tensor $\sigma^c_{ij}=\sigma^{c,\uparrow}_{ij}
+\sigma^{c,\downarrow}_{ij}$ is therefore diagonal, whereas the spin-current tensor $\sigma^s_{ij}=\frac{\hbar}{2e}\left(\sigma^{c,\uparrow}_{ij}-\sigma^{c,\downarrow}_{ij}\right)\equiv\frac{\hbar}{e}\tilde{\sigma}^s_{ij}$ is purely off-diagonal. Here $\tilde{\sigma}^s_{ij}$ is the charge-equivalent spin conductivity and has the same units as the charge conductivity. An electric field parallel or normal to the mirror plane thus produces a purely longitudinal charge current together with a purely transverse Drude spin current.

\begin{figure}[t]
	\centering
	\includegraphics[width=\columnwidth]{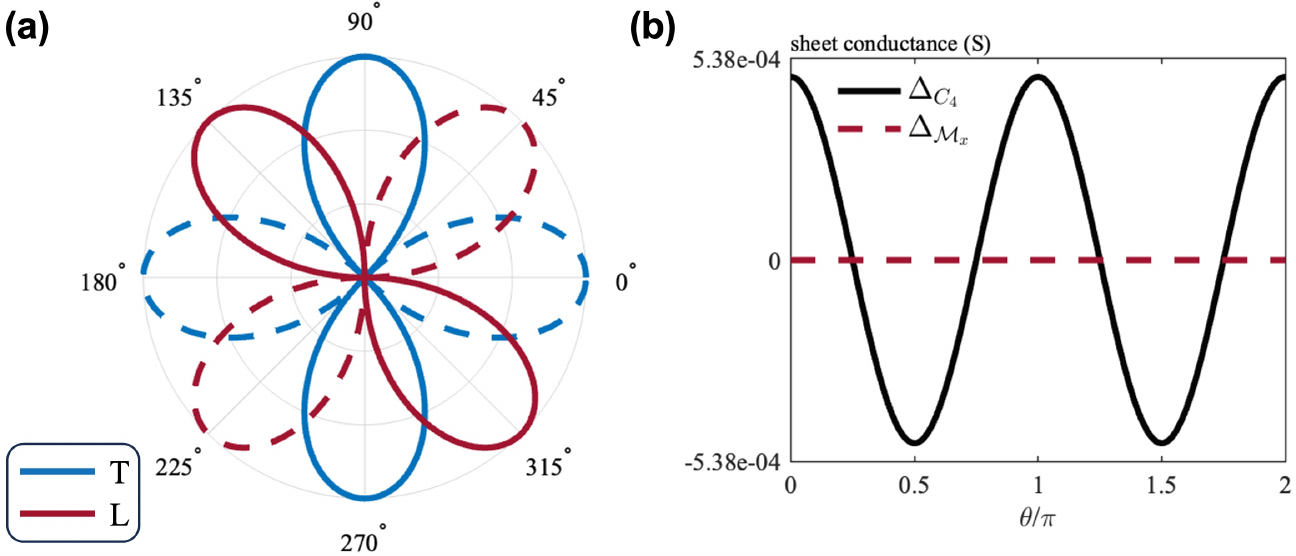}
	\caption{Drude spin-current response and symmetry probe for the mirror-governed stripe altermagnet. (a) Normalized longitudinal (L) and transverse (T) charge-equivalent spin conductances for $\bm E=E(\cos\theta,\sin\theta)$. Solid and dashed curves denote opposite signs. (b) Spin-resolved diagnostics: the $C_{4z}$-rotated difference is finite, while the $[C_2\Vert\mathcal{M}_x]$ mirror residual vanishes. Parameters: $t_{x0}=0.35~\mathrm{eV}$, $t_{x1}=0.07~\mathrm{eV}$, $t_{y0}=0.0525~\mathrm{eV}$, $t_{y1}=0.0105~\mathrm{eV}$, $t_{d0}=0.14~\mathrm{eV}$, $t_{d1}=0.105~\mathrm{eV}$, $J=0.14~\mathrm{eV}$, and $\delta_{x}=0.105~\mathrm{eV}$.}
	\label{fig4}
\end{figure}

For an in-plane field direction $\hat{\bm E}=(\cos\theta,\sin\theta)^{T}$, the longitudinal and transverse charge-equivalent spin responses per unit field are
\begin{align}
	\tilde{J}^s_L(\theta)&=\hat{\bm E}^{T}\tilde{\sigma}^s\hat{\bm E}
	=\tilde{\sigma}_{xy}^{s}\sin2\theta,\nonumber\\
	\tilde{J}^s_T(\theta)&=\hat{\bm E}_{\perp}^{T}\tilde{\sigma}^s\hat{\bm E}
	=\tilde{\sigma}_{xy}^{s}\cos2\theta,
	\label{eq:angular_drude_spin_current}
\end{align}
where $\hat{\bm E}_{\perp}=(-\sin\theta,\cos\theta)^{T}$. Both projections yield four-lobed polar patterns [Fig.~\ref{fig4}(a)]. This apparent fourfold structure was first predicted for Néel AM because of its inherent spin-reversing mirror symmetry~\cite{Ma2021AM}. However, spin-reversing mirror and rotation symmetries typically coexist in Néel AM, and the four-lobed polar pattern does not reveal whether the spin-reversing rotation is present. It therefore does not by itself distinguish mirror-governed nematic altermagnetism from a rotation-governed $l$-wave state.

The mirror- and rotation-based scenarios can instead be distinguished directly through the spin-resolved longitudinal charge conductivities. Defining $\sigma^{c,s}_L(\theta)=\hat{\bm E}^T\sigma^{c,s}\hat{\bm E}$, the rotation and mirror probes are
\begin{eqnarray}
	\Delta_{C_4}(\theta)&=&
	\sigma^{c,\uparrow}_L(\theta)-\sigma^{c,\downarrow}_L(\theta+\pi/2),
	\label{eq:c4_probe}\\
	\Delta_{\mathcal{M}_x}(\theta)&=&
	\sigma^{c,\uparrow}_L(\theta)-\sigma^{c,\downarrow}_L(\pi-\theta).
	\label{eq:mirror_probe}
\end{eqnarray}
The probes vanish when the spin sectors are related by $C_{4}$ and $\mathcal{M}_{x}$, respectively. Using Eq.~\eqref{eq:spin_resolved_tensor_form}, $\Delta_{C_4}(\theta)=(\sigma_{xx}^{c,\uparrow}-\sigma_{yy}^{c,\uparrow})\cos2\theta$, whereas $\Delta_{\mathcal{M}_x}(\theta)$ vanishes identically. The rotational probe extends directly to any candidate $[C_2\Vert C_{nz}]$ symmetry by replacing the angular shift $\pi/2$ in Eq.~\eqref{eq:c4_probe} with $2\pi/n$.

For the two-dimensional minimal orbital-order model, we use the representative parameters in Fig.~\ref{fig4}, together with $k_BT=0.0105~\mathrm{eV}$ and $\tau=10~\mathrm{fs}$. The maximal charge-equivalent spin sheet conductance is $|\tilde{\sigma}_{xy}^{s}|=2.522\times10^{-4}~\mathrm{S}$ at $\mu=0.196~\mathrm{eV}$~\cite{supplemental}, while the $C_{4z}$ probe in Fig.~\ref{fig4}(b) reaches $4.843\times10^{-4}~\mathrm{S}$ and the mirror probe vanishes by symmetry. These $10^{-4}$-S signals should be accessible to state-of-the-art spin-sensitive transport measurements~\cite{Bai2023AM}.

{\it \color{blue}Discussion and conclusions.--} We have introduced altermagnetism into stripe-order AFMs and established the resulting stripe-order altermagnetism as a realization of mirror-governed nematic altermagnetism beyond the rotation-based $l$-wave classification. A spin-reversing mirror relates the opposite-spin sectors and constrains their anisotropic spin-splitting pattern, while every crystallographic spin-reversing rotation is absent.
For the two-orbital parent model, RPA identifies the stripe-AM ordering channels near half filling under strong hopping anisotropy. For electric fields parallel or normal to the spin-reversing mirror plane, symmetry enforces a purely transverse Drude spin current together with a purely longitudinal charge current. Combined with the spin-resolved mirror and rotation probes, this response distinguishes mirror-governed nematic altermagnets from rotation-governed $l$-wave altermagnets.

Stripe magnetism has been predicted or observed in many correlated systems, providing potential platforms for stripe AM when the required orbital order and spin-reversing mirror symmetry are present. Examples include stripe-ordered parent compounds of iron-pnictide superconductors~\cite{Fernandes2012nematic}, the (Eu-doped) high-$T_c$ cuprate La$_{1.8-x}$Eu$_{0.2}$Sr$_x$CuO$_4$~\cite{Missiaen2025stripe}, the cobalt chalcogenide BaCoS$_2$~\cite{lenz2024}, and the perovskite manganite Sr$_{1-x}$Sm$_x$MnO$_3$~\cite{yamagata2017}. Of particular relevance, Sr$_{1-x}$Sm$_x$MnO$_3$ may host simultaneous stripe spin and orbital orders for $0.33\leq x\leq0.38$~\cite{yamagata2017}, whereas BaCoS$_2$ is predicted to combine stripe spin order with Néel orbital order within each $ab$ plane~\cite{lenz2024}.

This characteristic spin-reversing mirror also makes stripe AM a natural parent state for the recently proposed mixed-parity altermagnetism~\cite{zhuang2026mixed}. In a mixed-parity altermagnet, the spin-polarized bands satisfy neither the even-parity relation $E_{\bk,s}=E_{-\bk,s}$ nor the odd-parity relation $E_{\bk,s}=E_{-\bk,-s}$; nevertheless, a spin-reversing mirror can enforce zero net spin magnetization and nematic spin splitting. By contrast, stripe AM preserves the spinless time-reversal symmetry $[\bar{C}_{2}\Vert\mathcal{T}]$ and therefore exhibits even-parity spin splitting. In a noncentrosymmetric stripe AM, applying circularly polarized light can break $[\bar{C}_{2}\Vert\mathcal{T}]$ and provide a route from even- to mixed-parity spin splitting.

In summary, stripe AMs are distinguished from Néel AMs not only by their real-space magnetic order but also by their mirror-constrained nematic spin splitting and symmetry-constrained spin-current responses. Together, these results establish stripe AMs as a natural form of nematic altermagnetism---a mirror-based class beyond the conventional rotation-based $l$-wave framework.

{\it \color{blue}Acknowledgments.--} Z.-Y. Zhuang thanks Lun-Hui Hu for fruitful discussions on a closely related project. Z.-Y. Zhuang and Z. Yan acknowledge support from the Fundamental and Interdisciplinary Disciplines Breakthrough Plan of the Ministry of Education of China (Grant No. JYB2025XDXM403) and the Guangdong Basic and Applied Basic Research Foundation (Grant No. 2023B1515040023).

\bibliography{dirac.bib}

\begin{thebibliography}{83}%
\makeatletter
\providecommand \@ifxundefined [1]{%
 \@ifx{#1\undefined}
}%
\providecommand \@ifnum [1]{%
 \ifnum #1\expandafter \@firstoftwo
 \else \expandafter \@secondoftwo
 \fi
}%
\providecommand \@ifx [1]{%
 \ifx #1\expandafter \@firstoftwo
 \else \expandafter \@secondoftwo
 \fi
}%
\providecommand \natexlab [1]{#1}%
\providecommand \enquote  [1]{``#1''}%
\providecommand \bibnamefont  [1]{#1}%
\providecommand \bibfnamefont [1]{#1}%
\providecommand \citenamefont [1]{#1}%
\providecommand \href@noop [0]{\@secondoftwo}%
\providecommand \href [0]{\begingroup \@sanitize@url \@href}%
\providecommand \@href[1]{\@@startlink{#1}\@@href}%
\providecommand \@@href[1]{\endgroup#1\@@endlink}%
\providecommand \@sanitize@url [0]{\catcode `\\12\catcode `\$12\catcode
  `\&12\catcode `\#12\catcode `\^12\catcode `\_12\catcode `\%12\relax}%
\providecommand \@@startlink[1]{}%
\providecommand \@@endlink[0]{}%
\providecommand \url  [0]{\begingroup\@sanitize@url \@url }%
\providecommand \@url [1]{\endgroup\@href {#1}{\urlprefix }}%
\providecommand \urlprefix  [0]{URL }%
\providecommand \Eprint [0]{\href }%
\providecommand \doibase [0]{https://doi.org/}%
\providecommand \selectlanguage [0]{\@gobble}%
\providecommand \bibinfo  [0]{\@secondoftwo}%
\providecommand \bibfield  [0]{\@secondoftwo}%
\providecommand \translation [1]{[#1]}%
\providecommand \BibitemOpen [0]{}%
\providecommand \bibitemStop [0]{}%
\providecommand \bibitemNoStop [0]{.\EOS\space}%
\providecommand \EOS [0]{\spacefactor3000\relax}%
\providecommand \BibitemShut  [1]{\csname bibitem#1\endcsname}%
\let\auto@bib@innerbib\@empty
\bibitem [{\citenamefont {Hayami}\ \emph {et~al.}(2019)\citenamefont {Hayami},
  \citenamefont {Yanagi},\ and\ \citenamefont {Kusunose}}]{Hayami2019NRSS}%
  \BibitemOpen
  \bibfield  {author} {\bibinfo {author} {\bibfnamefont {S.}~\bibnamefont
  {Hayami}}, \bibinfo {author} {\bibfnamefont {Y.}~\bibnamefont {Yanagi}},\
  and\ \bibinfo {author} {\bibfnamefont {H.}~\bibnamefont {Kusunose}},\
  }\bibfield  {title} {\bibinfo {title} {Momentum-dependent spin splitting by
  collinear antiferromagnetic ordering},\ }\href
  {https://doi.org/10.7566/JPSJ.88.123702} {\bibfield  {journal} {\bibinfo
  {journal} {Journal of the Physical Society of Japan}\ }\textbf {\bibinfo
  {volume} {88}},\ \bibinfo {pages} {123702} (\bibinfo {year}
  {2019})}\BibitemShut {NoStop}%
\bibitem [{\citenamefont {Hayami}\ \emph {et~al.}(2020)\citenamefont {Hayami},
  \citenamefont {Yanagi},\ and\ \citenamefont {Kusunose}}]{Hayami2020NRSS}%
  \BibitemOpen
  \bibfield  {author} {\bibinfo {author} {\bibfnamefont {S.}~\bibnamefont
  {Hayami}}, \bibinfo {author} {\bibfnamefont {Y.}~\bibnamefont {Yanagi}},\
  and\ \bibinfo {author} {\bibfnamefont {H.}~\bibnamefont {Kusunose}},\
  }\bibfield  {title} {\bibinfo {title} {Bottom-up design of spin-split and
  reshaped electronic band structures in antiferromagnets without spin-orbit
  coupling: Procedure on the basis of augmented multipoles},\ }\href
  {https://doi.org/10.1103/PhysRevB.102.144441} {\bibfield  {journal} {\bibinfo
   {journal} {Phys. Rev. B}\ }\textbf {\bibinfo {volume} {102}},\ \bibinfo
  {pages} {144441} (\bibinfo {year} {2020})}\BibitemShut {NoStop}%
\bibitem [{\citenamefont {Liu}\ \emph {et~al.}(2025)\citenamefont {Liu},
  \citenamefont {Dai},\ and\ \citenamefont {Blügel}}]{Liu2025}%
  \BibitemOpen
  \bibfield  {author} {\bibinfo {author} {\bibfnamefont {Q.}~\bibnamefont
  {Liu}}, \bibinfo {author} {\bibfnamefont {X.}~\bibnamefont {Dai}},\ and\
  \bibinfo {author} {\bibfnamefont {S.}~\bibnamefont {Blügel}},\ }\bibfield
  {title} {\bibinfo {title} {Different facets of unconventional magnetism},\
  }\href {https://doi.org/10.1038/s41567-024-02750-3} {\bibfield  {journal}
  {\bibinfo  {journal} {Nature Physics}\ }\textbf {\bibinfo {volume} {21}},\
  \bibinfo {pages} {329} (\bibinfo {year} {2025})}\BibitemShut {NoStop}%
\bibitem [{\citenamefont {Ma}\ \emph {et~al.}(2021)\citenamefont {Ma},
  \citenamefont {Hu}, \citenamefont {Li}, \citenamefont {Liu}, \citenamefont
  {Yao}, \citenamefont {Jia},\ and\ \citenamefont {Liu}}]{Ma2021AM}%
  \BibitemOpen
  \bibfield  {author} {\bibinfo {author} {\bibfnamefont {H.-Y.}\ \bibnamefont
  {Ma}}, \bibinfo {author} {\bibfnamefont {M.}~\bibnamefont {Hu}}, \bibinfo
  {author} {\bibfnamefont {N.}~\bibnamefont {Li}}, \bibinfo {author}
  {\bibfnamefont {J.}~\bibnamefont {Liu}}, \bibinfo {author} {\bibfnamefont
  {W.}~\bibnamefont {Yao}}, \bibinfo {author} {\bibfnamefont {J.-F.}\
  \bibnamefont {Jia}},\ and\ \bibinfo {author} {\bibfnamefont {J.}~\bibnamefont
  {Liu}},\ }\bibfield  {title} {\bibinfo {title} {Multifunctional
  antiferromagnetic materials with giant piezomagnetism and noncollinear spin
  current},\ }\href {https://doi.org/10.1038/s41467-021-23127-7} {\bibfield
  {journal} {\bibinfo  {journal} {Nature Communications}\ }\textbf {\bibinfo
  {volume} {12}},\ \bibinfo {pages} {2846} (\bibinfo {year}
  {2021})}\BibitemShut {NoStop}%
\bibitem [{\citenamefont {\ifmmode~\check{S}\else \v{S}\fi{}mejkal}\ \emph
  {et~al.}(2022{\natexlab{a}})\citenamefont {\ifmmode~\check{S}\else
  \v{S}\fi{}mejkal}, \citenamefont {Sinova},\ and\ \citenamefont
  {Jungwirth}}]{Libor2022AMa}%
  \BibitemOpen
  \bibfield  {author} {\bibinfo {author} {\bibfnamefont {L.}~\bibnamefont
  {\ifmmode~\check{S}\else \v{S}\fi{}mejkal}}, \bibinfo {author} {\bibfnamefont
  {J.}~\bibnamefont {Sinova}},\ and\ \bibinfo {author} {\bibfnamefont
  {T.}~\bibnamefont {Jungwirth}},\ }\bibfield  {title} {\bibinfo {title}
  {{Beyond Conventional Ferromagnetism and Antiferromagnetism: A Phase with
  Nonrelativistic Spin and Crystal Rotation Symmetry}},\ }\href
  {https://doi.org/10.1103/PhysRevX.12.031042} {\bibfield  {journal} {\bibinfo
  {journal} {Phys. Rev. X}\ }\textbf {\bibinfo {volume} {12}},\ \bibinfo
  {pages} {031042} (\bibinfo {year} {2022}{\natexlab{a}})}\BibitemShut
  {NoStop}%
\bibitem [{\citenamefont {Yuan}\ \emph {et~al.}(2020)\citenamefont {Yuan},
  \citenamefont {Wang}, \citenamefont {Luo}, \citenamefont {Rashba},\ and\
  \citenamefont {Zunger}}]{Yuan2020AM}%
  \BibitemOpen
  \bibfield  {author} {\bibinfo {author} {\bibfnamefont {L.-D.}\ \bibnamefont
  {Yuan}}, \bibinfo {author} {\bibfnamefont {Z.}~\bibnamefont {Wang}}, \bibinfo
  {author} {\bibfnamefont {J.-W.}\ \bibnamefont {Luo}}, \bibinfo {author}
  {\bibfnamefont {E.~I.}\ \bibnamefont {Rashba}},\ and\ \bibinfo {author}
  {\bibfnamefont {A.}~\bibnamefont {Zunger}},\ }\bibfield  {title} {\bibinfo
  {title} {{Giant momentum-dependent spin splitting in centrosymmetric low-$Z$
  antiferromagnets}},\ }\href {https://doi.org/10.1103/PhysRevB.102.014422}
  {\bibfield  {journal} {\bibinfo  {journal} {Phys. Rev. B}\ }\textbf {\bibinfo
  {volume} {102}},\ \bibinfo {pages} {014422} (\bibinfo {year}
  {2020})}\BibitemShut {NoStop}%
\bibitem [{\citenamefont {Yuan}\ \emph {et~al.}(2021)\citenamefont {Yuan},
  \citenamefont {Wang}, \citenamefont {Luo},\ and\ \citenamefont
  {Zunger}}]{Yuan2021AM}%
  \BibitemOpen
  \bibfield  {author} {\bibinfo {author} {\bibfnamefont {L.-D.}\ \bibnamefont
  {Yuan}}, \bibinfo {author} {\bibfnamefont {Z.}~\bibnamefont {Wang}}, \bibinfo
  {author} {\bibfnamefont {J.-W.}\ \bibnamefont {Luo}},\ and\ \bibinfo {author}
  {\bibfnamefont {A.}~\bibnamefont {Zunger}},\ }\bibfield  {title} {\bibinfo
  {title} {{Prediction of low-Z collinear and noncollinear antiferromagnetic
  compounds having momentum-dependent spin splitting even without spin-orbit
  coupling}},\ }\href {https://doi.org/10.1103/PhysRevMaterials.5.014409}
  {\bibfield  {journal} {\bibinfo  {journal} {Phys. Rev. Mater.}\ }\textbf
  {\bibinfo {volume} {5}},\ \bibinfo {pages} {014409} (\bibinfo {year}
  {2021})}\BibitemShut {NoStop}%
\bibitem [{\citenamefont {Hirsch}(1990)}]{Hirsch1990}%
  \BibitemOpen
  \bibfield  {author} {\bibinfo {author} {\bibfnamefont {J.~E.}\ \bibnamefont
  {Hirsch}},\ }\bibfield  {title} {\bibinfo {title} {Spin-split states in
  metals},\ }\href {https://doi.org/10.1103/PhysRevB.41.6820} {\bibfield
  {journal} {\bibinfo  {journal} {Phys. Rev. B}\ }\textbf {\bibinfo {volume}
  {41}},\ \bibinfo {pages} {6820} (\bibinfo {year} {1990})}\BibitemShut
  {NoStop}%
\bibitem [{\citenamefont {Ikeda}\ and\ \citenamefont
  {Ohashi}(1998)}]{Ikeda1998}%
  \BibitemOpen
  \bibfield  {author} {\bibinfo {author} {\bibfnamefont {H.}~\bibnamefont
  {Ikeda}}\ and\ \bibinfo {author} {\bibfnamefont {Y.}~\bibnamefont {Ohashi}},\
  }\bibfield  {title} {\bibinfo {title} {Theory of unconventional spin density
  wave: A possible mechanism of the micromagnetism in u-based heavy fermion
  compounds},\ }\href {https://doi.org/10.1103/PhysRevLett.81.3723} {\bibfield
  {journal} {\bibinfo  {journal} {Phys. Rev. Lett.}\ }\textbf {\bibinfo
  {volume} {81}},\ \bibinfo {pages} {3723} (\bibinfo {year}
  {1998})}\BibitemShut {NoStop}%
\bibitem [{\citenamefont {Wu}\ and\ \citenamefont {Zhang}(2004)}]{wu2004}%
  \BibitemOpen
  \bibfield  {author} {\bibinfo {author} {\bibfnamefont {C.}~\bibnamefont
  {Wu}}\ and\ \bibinfo {author} {\bibfnamefont {S.-C.}\ \bibnamefont {Zhang}},\
  }\bibfield  {title} {\bibinfo {title} {Dynamic generation of spin-orbit
  coupling},\ }\href {https://doi.org/10.1103/PhysRevLett.93.036403} {\bibfield
   {journal} {\bibinfo  {journal} {Phys. Rev. Lett.}\ }\textbf {\bibinfo
  {volume} {93}},\ \bibinfo {pages} {036403} (\bibinfo {year}
  {2004})}\BibitemShut {NoStop}%
\bibitem [{\citenamefont {Wu}\ \emph {et~al.}(2007)\citenamefont {Wu},
  \citenamefont {Sun}, \citenamefont {Fradkin},\ and\ \citenamefont
  {Zhang}}]{wu2007}%
  \BibitemOpen
  \bibfield  {author} {\bibinfo {author} {\bibfnamefont {C.}~\bibnamefont
  {Wu}}, \bibinfo {author} {\bibfnamefont {K.}~\bibnamefont {Sun}}, \bibinfo
  {author} {\bibfnamefont {E.}~\bibnamefont {Fradkin}},\ and\ \bibinfo {author}
  {\bibfnamefont {S.-C.}\ \bibnamefont {Zhang}},\ }\bibfield  {title} {\bibinfo
  {title} {{Fermi liquid instabilities in the spin channel}},\ }\href
  {https://doi.org/10.1103/PhysRevB.75.115103} {\bibfield  {journal} {\bibinfo
  {journal} {Phys. Rev. B}\ }\textbf {\bibinfo {volume} {75}},\ \bibinfo
  {pages} {115103} (\bibinfo {year} {2007})}\BibitemShut {NoStop}%
\bibitem [{\citenamefont {Reimers}\ \emph {et~al.}(2024)\citenamefont
  {Reimers}, \citenamefont {Odenbreit}, \citenamefont {{\v{S}}mejkal},
  \citenamefont {Strocov}, \citenamefont {Constantinou}, \citenamefont
  {Hellenes}, \citenamefont {Jaeschke~Ubiergo}, \citenamefont {Campos},
  \citenamefont {Bharadwaj}, \citenamefont {Chakraborty}, \citenamefont
  {Denneulin}, \citenamefont {Shi}, \citenamefont {Dunin-Borkowski},
  \citenamefont {Das}, \citenamefont {Kl{\"a}ui}, \citenamefont {Sinova},\ and\
  \citenamefont {Jourdan}}]{Reimers2024}%
  \BibitemOpen
  \bibfield  {author} {\bibinfo {author} {\bibfnamefont {S.}~\bibnamefont
  {Reimers}}, \bibinfo {author} {\bibfnamefont {L.}~\bibnamefont {Odenbreit}},
  \bibinfo {author} {\bibfnamefont {L.}~\bibnamefont {{\v{S}}mejkal}}, \bibinfo
  {author} {\bibfnamefont {V.~N.}\ \bibnamefont {Strocov}}, \bibinfo {author}
  {\bibfnamefont {P.}~\bibnamefont {Constantinou}}, \bibinfo {author}
  {\bibfnamefont {A.~B.}\ \bibnamefont {Hellenes}}, \bibinfo {author}
  {\bibfnamefont {R.}~\bibnamefont {Jaeschke~Ubiergo}}, \bibinfo {author}
  {\bibfnamefont {W.~H.}\ \bibnamefont {Campos}}, \bibinfo {author}
  {\bibfnamefont {V.~K.}\ \bibnamefont {Bharadwaj}}, \bibinfo {author}
  {\bibfnamefont {A.}~\bibnamefont {Chakraborty}}, \bibinfo {author}
  {\bibfnamefont {T.}~\bibnamefont {Denneulin}}, \bibinfo {author}
  {\bibfnamefont {W.}~\bibnamefont {Shi}}, \bibinfo {author} {\bibfnamefont
  {R.~E.}\ \bibnamefont {Dunin-Borkowski}}, \bibinfo {author} {\bibfnamefont
  {S.}~\bibnamefont {Das}}, \bibinfo {author} {\bibfnamefont {M.}~\bibnamefont
  {Kl{\"a}ui}}, \bibinfo {author} {\bibfnamefont {J.}~\bibnamefont {Sinova}},\
  and\ \bibinfo {author} {\bibfnamefont {M.}~\bibnamefont {Jourdan}},\
  }\bibfield  {title} {\bibinfo {title} {{Direct observation of altermagnetic
  band splitting in CrSb thin films}},\ }\href
  {https://doi.org/10.1038/s41467-024-46476-5} {\bibfield  {journal} {\bibinfo
  {journal} {Nature Communications}\ }\textbf {\bibinfo {volume} {15}},\
  \bibinfo {pages} {2116} (\bibinfo {year} {2024})}\BibitemShut {NoStop}%
\bibitem [{\citenamefont {Ding}\ \emph {et~al.}(2024)\citenamefont {Ding},
  \citenamefont {Jiang}, \citenamefont {Chen}, \citenamefont {Tao},
  \citenamefont {Liu}, \citenamefont {Li}, \citenamefont {Liu}, \citenamefont
  {Sun}, \citenamefont {Cheng}, \citenamefont {Liu}, \citenamefont {Yang},
  \citenamefont {Zhang}, \citenamefont {Deng}, \citenamefont {Jing},
  \citenamefont {Huang}, \citenamefont {Shi}, \citenamefont {Ye}, \citenamefont
  {Qiao}, \citenamefont {Wang}, \citenamefont {Guo}, \citenamefont {Feng},\
  and\ \citenamefont {Shen}}]{Ding2024CrSb}%
  \BibitemOpen
  \bibfield  {author} {\bibinfo {author} {\bibfnamefont {J.}~\bibnamefont
  {Ding}}, \bibinfo {author} {\bibfnamefont {Z.}~\bibnamefont {Jiang}},
  \bibinfo {author} {\bibfnamefont {X.}~\bibnamefont {Chen}}, \bibinfo {author}
  {\bibfnamefont {Z.}~\bibnamefont {Tao}}, \bibinfo {author} {\bibfnamefont
  {Z.}~\bibnamefont {Liu}}, \bibinfo {author} {\bibfnamefont {T.}~\bibnamefont
  {Li}}, \bibinfo {author} {\bibfnamefont {J.}~\bibnamefont {Liu}}, \bibinfo
  {author} {\bibfnamefont {J.}~\bibnamefont {Sun}}, \bibinfo {author}
  {\bibfnamefont {J.}~\bibnamefont {Cheng}}, \bibinfo {author} {\bibfnamefont
  {J.}~\bibnamefont {Liu}}, \bibinfo {author} {\bibfnamefont {Y.}~\bibnamefont
  {Yang}}, \bibinfo {author} {\bibfnamefont {R.}~\bibnamefont {Zhang}},
  \bibinfo {author} {\bibfnamefont {L.}~\bibnamefont {Deng}}, \bibinfo {author}
  {\bibfnamefont {W.}~\bibnamefont {Jing}}, \bibinfo {author} {\bibfnamefont
  {Y.}~\bibnamefont {Huang}}, \bibinfo {author} {\bibfnamefont
  {Y.}~\bibnamefont {Shi}}, \bibinfo {author} {\bibfnamefont {M.}~\bibnamefont
  {Ye}}, \bibinfo {author} {\bibfnamefont {S.}~\bibnamefont {Qiao}}, \bibinfo
  {author} {\bibfnamefont {Y.}~\bibnamefont {Wang}}, \bibinfo {author}
  {\bibfnamefont {Y.}~\bibnamefont {Guo}}, \bibinfo {author} {\bibfnamefont
  {D.}~\bibnamefont {Feng}},\ and\ \bibinfo {author} {\bibfnamefont
  {D.}~\bibnamefont {Shen}},\ }\bibfield  {title} {\bibinfo {title} {{Large
  Band Splitting in $g$-Wave Altermagnet CrSb}},\ }\href
  {https://doi.org/10.1103/PhysRevLett.133.206401} {\bibfield  {journal}
  {\bibinfo  {journal} {Phys. Rev. Lett.}\ }\textbf {\bibinfo {volume} {133}},\
  \bibinfo {pages} {206401} (\bibinfo {year} {2024})}\BibitemShut {NoStop}%
\bibitem [{\citenamefont {Yang}\ \emph {et~al.}(2025)\citenamefont {Yang},
  \citenamefont {Li}, \citenamefont {Yang}, \citenamefont {Li}, \citenamefont
  {Zheng}, \citenamefont {Zhu}, \citenamefont {Pan}, \citenamefont {Xu},
  \citenamefont {Cao}, \citenamefont {Zhao}, \citenamefont {Jana},
  \citenamefont {Zhang}, \citenamefont {Ye}, \citenamefont {Song},
  \citenamefont {Hu}, \citenamefont {Yang}, \citenamefont {Fujii},
  \citenamefont {Vobornik}, \citenamefont {Shi}, \citenamefont {Yuan},
  \citenamefont {Zhang}, \citenamefont {Xu},\ and\ \citenamefont
  {Liu}}]{Yang2024CrSb}%
  \BibitemOpen
  \bibfield  {author} {\bibinfo {author} {\bibfnamefont {G.}~\bibnamefont
  {Yang}}, \bibinfo {author} {\bibfnamefont {Z.}~\bibnamefont {Li}}, \bibinfo
  {author} {\bibfnamefont {S.}~\bibnamefont {Yang}}, \bibinfo {author}
  {\bibfnamefont {J.}~\bibnamefont {Li}}, \bibinfo {author} {\bibfnamefont
  {H.}~\bibnamefont {Zheng}}, \bibinfo {author} {\bibfnamefont
  {W.}~\bibnamefont {Zhu}}, \bibinfo {author} {\bibfnamefont {Z.}~\bibnamefont
  {Pan}}, \bibinfo {author} {\bibfnamefont {Y.}~\bibnamefont {Xu}}, \bibinfo
  {author} {\bibfnamefont {S.}~\bibnamefont {Cao}}, \bibinfo {author}
  {\bibfnamefont {W.}~\bibnamefont {Zhao}}, \bibinfo {author} {\bibfnamefont
  {A.}~\bibnamefont {Jana}}, \bibinfo {author} {\bibfnamefont {J.}~\bibnamefont
  {Zhang}}, \bibinfo {author} {\bibfnamefont {M.}~\bibnamefont {Ye}}, \bibinfo
  {author} {\bibfnamefont {Y.}~\bibnamefont {Song}}, \bibinfo {author}
  {\bibfnamefont {L.-H.}\ \bibnamefont {Hu}}, \bibinfo {author} {\bibfnamefont
  {L.}~\bibnamefont {Yang}}, \bibinfo {author} {\bibfnamefont {J.}~\bibnamefont
  {Fujii}}, \bibinfo {author} {\bibfnamefont {I.}~\bibnamefont {Vobornik}},
  \bibinfo {author} {\bibfnamefont {M.}~\bibnamefont {Shi}}, \bibinfo {author}
  {\bibfnamefont {H.}~\bibnamefont {Yuan}}, \bibinfo {author} {\bibfnamefont
  {Y.}~\bibnamefont {Zhang}}, \bibinfo {author} {\bibfnamefont
  {Y.}~\bibnamefont {Xu}},\ and\ \bibinfo {author} {\bibfnamefont
  {Y.}~\bibnamefont {Liu}},\ }\bibfield  {title} {\bibinfo {title}
  {{Three-dimensional mapping of the altermagnetic spin splitting in CrSb}},\
  }\href {https://doi.org/10.1038/s41467-025-56647-7} {\bibfield  {journal}
  {\bibinfo  {journal} {Nature Communications}\ }\textbf {\bibinfo {volume}
  {16}},\ \bibinfo {pages} {1442} (\bibinfo {year} {2025})}\BibitemShut
  {NoStop}%
\bibitem [{\citenamefont {Zeng}\ \emph {et~al.}(2024)\citenamefont {Zeng},
  \citenamefont {Zhu}, \citenamefont {Zhu}, \citenamefont {Liu}, \citenamefont
  {Ma}, \citenamefont {Hao}, \citenamefont {Liu}, \citenamefont {Qu},
  \citenamefont {Yang}, \citenamefont {Jiang}, \citenamefont {Yamagami},
  \citenamefont {Arita}, \citenamefont {Zhang}, \citenamefont {Shao},
  \citenamefont {Dai}, \citenamefont {Shimada}, \citenamefont {Liu},
  \citenamefont {Ye}, \citenamefont {Huang}, \citenamefont {Liu},\ and\
  \citenamefont {Liu}}]{Zeng2024CrSb}%
  \BibitemOpen
  \bibfield  {author} {\bibinfo {author} {\bibfnamefont {M.}~\bibnamefont
  {Zeng}}, \bibinfo {author} {\bibfnamefont {M.-Y.}\ \bibnamefont {Zhu}},
  \bibinfo {author} {\bibfnamefont {Y.-P.}\ \bibnamefont {Zhu}}, \bibinfo
  {author} {\bibfnamefont {X.-R.}\ \bibnamefont {Liu}}, \bibinfo {author}
  {\bibfnamefont {X.-M.}\ \bibnamefont {Ma}}, \bibinfo {author} {\bibfnamefont
  {Y.-J.}\ \bibnamefont {Hao}}, \bibinfo {author} {\bibfnamefont
  {P.}~\bibnamefont {Liu}}, \bibinfo {author} {\bibfnamefont {G.}~\bibnamefont
  {Qu}}, \bibinfo {author} {\bibfnamefont {Y.}~\bibnamefont {Yang}}, \bibinfo
  {author} {\bibfnamefont {Z.}~\bibnamefont {Jiang}}, \bibinfo {author}
  {\bibfnamefont {K.}~\bibnamefont {Yamagami}}, \bibinfo {author}
  {\bibfnamefont {M.}~\bibnamefont {Arita}}, \bibinfo {author} {\bibfnamefont
  {X.}~\bibnamefont {Zhang}}, \bibinfo {author} {\bibfnamefont {T.-H.}\
  \bibnamefont {Shao}}, \bibinfo {author} {\bibfnamefont {Y.}~\bibnamefont
  {Dai}}, \bibinfo {author} {\bibfnamefont {K.}~\bibnamefont {Shimada}},
  \bibinfo {author} {\bibfnamefont {Z.}~\bibnamefont {Liu}}, \bibinfo {author}
  {\bibfnamefont {M.}~\bibnamefont {Ye}}, \bibinfo {author} {\bibfnamefont
  {Y.}~\bibnamefont {Huang}}, \bibinfo {author} {\bibfnamefont
  {Q.}~\bibnamefont {Liu}},\ and\ \bibinfo {author} {\bibfnamefont
  {C.}~\bibnamefont {Liu}},\ }\bibfield  {title} {\bibinfo {title}
  {{Observation of Spin Splitting in Room-Temperature Metallic Antiferromagnet
  CrSb}},\ }\href {https://doi.org/https://doi.org/10.1002/advs.202406529}
  {\bibfield  {journal} {\bibinfo  {journal} {Advanced Science}\ }\textbf
  {\bibinfo {volume} {11}},\ \bibinfo {pages} {2406529} (\bibinfo {year}
  {2024})}\BibitemShut {NoStop}%
\bibitem [{\citenamefont {Li}\ \emph {et~al.}(2025)\citenamefont {Li},
  \citenamefont {Hu}, \citenamefont {Li}, \citenamefont {Wang}, \citenamefont
  {Chen}, \citenamefont {Thiagarajan}, \citenamefont {Leandersson},
  \citenamefont {Polley}, \citenamefont {Kim}, \citenamefont {Liu},
  \citenamefont {Fulga}, \citenamefont {Vergniory}, \citenamefont {Janson},
  \citenamefont {Tjernberg},\ and\ \citenamefont {van~den Brink}}]{Li2024CrSb}%
  \BibitemOpen
  \bibfield  {author} {\bibinfo {author} {\bibfnamefont {C.}~\bibnamefont
  {Li}}, \bibinfo {author} {\bibfnamefont {M.}~\bibnamefont {Hu}}, \bibinfo
  {author} {\bibfnamefont {Z.}~\bibnamefont {Li}}, \bibinfo {author}
  {\bibfnamefont {Y.}~\bibnamefont {Wang}}, \bibinfo {author} {\bibfnamefont
  {W.}~\bibnamefont {Chen}}, \bibinfo {author} {\bibfnamefont {B.}~\bibnamefont
  {Thiagarajan}}, \bibinfo {author} {\bibfnamefont {M.}~\bibnamefont
  {Leandersson}}, \bibinfo {author} {\bibfnamefont {C.}~\bibnamefont {Polley}},
  \bibinfo {author} {\bibfnamefont {T.}~\bibnamefont {Kim}}, \bibinfo {author}
  {\bibfnamefont {H.}~\bibnamefont {Liu}}, \bibinfo {author} {\bibfnamefont
  {C.}~\bibnamefont {Fulga}}, \bibinfo {author} {\bibfnamefont {M.~G.}\
  \bibnamefont {Vergniory}}, \bibinfo {author} {\bibfnamefont {O.}~\bibnamefont
  {Janson}}, \bibinfo {author} {\bibfnamefont {O.}~\bibnamefont {Tjernberg}},\
  and\ \bibinfo {author} {\bibfnamefont {J.}~\bibnamefont {van~den Brink}},\
  }\bibfield  {title} {\bibinfo {title} {{Topological Weyl altermagnetism in
  CrSb}},\ }\href {https://doi.org/10.1038/s42005-025-02232-9} {\bibfield
  {journal} {\bibinfo  {journal} {Communications Physics}\ }\textbf {\bibinfo
  {volume} {8}},\ \bibinfo {pages} {311} (\bibinfo {year} {2025})}\BibitemShut
  {NoStop}%
\bibitem [{\citenamefont {Lu}\ \emph {et~al.}(2025)\citenamefont {Lu},
  \citenamefont {Feng}, \citenamefont {Wang}, \citenamefont {Chen},
  \citenamefont {Lin}, \citenamefont {Liang}, \citenamefont {Liu},
  \citenamefont {Feng}, \citenamefont {Yamagami}, \citenamefont {Liu},
  \citenamefont {Felser}, \citenamefont {Wu},\ and\ \citenamefont
  {Ma}}]{Lu2024AM7}%
  \BibitemOpen
  \bibfield  {author} {\bibinfo {author} {\bibfnamefont {W.}~\bibnamefont
  {Lu}}, \bibinfo {author} {\bibfnamefont {S.}~\bibnamefont {Feng}}, \bibinfo
  {author} {\bibfnamefont {Y.}~\bibnamefont {Wang}}, \bibinfo {author}
  {\bibfnamefont {D.}~\bibnamefont {Chen}}, \bibinfo {author} {\bibfnamefont
  {Z.}~\bibnamefont {Lin}}, \bibinfo {author} {\bibfnamefont {X.}~\bibnamefont
  {Liang}}, \bibinfo {author} {\bibfnamefont {S.}~\bibnamefont {Liu}}, \bibinfo
  {author} {\bibfnamefont {W.}~\bibnamefont {Feng}}, \bibinfo {author}
  {\bibfnamefont {K.}~\bibnamefont {Yamagami}}, \bibinfo {author}
  {\bibfnamefont {J.}~\bibnamefont {Liu}}, \bibinfo {author} {\bibfnamefont
  {C.}~\bibnamefont {Felser}}, \bibinfo {author} {\bibfnamefont
  {Q.}~\bibnamefont {Wu}},\ and\ \bibinfo {author} {\bibfnamefont
  {J.}~\bibnamefont {Ma}},\ }\bibfield  {title} {\bibinfo {title} {{Signature
  of Topological Surface Bands in Altermagnetic Weyl Semimetal CrSb}},\ }\href
  {https://doi.org/10.1021/acs.nanolett.5c00482} {\bibfield  {journal}
  {\bibinfo  {journal} {Nano Letters}\ }\textbf {\bibinfo {volume} {25}},\
  \bibinfo {pages} {7343} (\bibinfo {year} {2025})}\BibitemShut {NoStop}%
\bibitem [{\citenamefont {Osumi}\ \emph {et~al.}(2024)\citenamefont {Osumi},
  \citenamefont {Souma}, \citenamefont {Aoyama}, \citenamefont {Yamauchi},
  \citenamefont {Honma}, \citenamefont {Nakayama}, \citenamefont {Takahashi},
  \citenamefont {Ohgushi},\ and\ \citenamefont {Sato}}]{Osumi2024MnTe}%
  \BibitemOpen
  \bibfield  {author} {\bibinfo {author} {\bibfnamefont {T.}~\bibnamefont
  {Osumi}}, \bibinfo {author} {\bibfnamefont {S.}~\bibnamefont {Souma}},
  \bibinfo {author} {\bibfnamefont {T.}~\bibnamefont {Aoyama}}, \bibinfo
  {author} {\bibfnamefont {K.}~\bibnamefont {Yamauchi}}, \bibinfo {author}
  {\bibfnamefont {A.}~\bibnamefont {Honma}}, \bibinfo {author} {\bibfnamefont
  {K.}~\bibnamefont {Nakayama}}, \bibinfo {author} {\bibfnamefont
  {T.}~\bibnamefont {Takahashi}}, \bibinfo {author} {\bibfnamefont
  {K.}~\bibnamefont {Ohgushi}},\ and\ \bibinfo {author} {\bibfnamefont
  {T.}~\bibnamefont {Sato}},\ }\bibfield  {title} {\bibinfo {title}
  {{Observation of a giant band splitting in altermagnetic MnTe}},\ }\href
  {https://doi.org/10.1103/PhysRevB.109.115102} {\bibfield  {journal} {\bibinfo
   {journal} {Phys. Rev. B}\ }\textbf {\bibinfo {volume} {109}},\ \bibinfo
  {pages} {115102} (\bibinfo {year} {2024})}\BibitemShut {NoStop}%
\bibitem [{\citenamefont {Lee}\ \emph {et~al.}(2024)\citenamefont {Lee},
  \citenamefont {Lee}, \citenamefont {Jung}, \citenamefont {Jung},
  \citenamefont {Kim}, \citenamefont {Lee}, \citenamefont {Seok}, \citenamefont
  {Kim}, \citenamefont {Park}, \citenamefont {\ifmmode~\check{S}\else
  \v{S}\fi{}mejkal}, \citenamefont {Kang},\ and\ \citenamefont
  {Kim}}]{Lee2024MnTe}%
  \BibitemOpen
  \bibfield  {author} {\bibinfo {author} {\bibfnamefont {S.}~\bibnamefont
  {Lee}}, \bibinfo {author} {\bibfnamefont {S.}~\bibnamefont {Lee}}, \bibinfo
  {author} {\bibfnamefont {S.}~\bibnamefont {Jung}}, \bibinfo {author}
  {\bibfnamefont {J.}~\bibnamefont {Jung}}, \bibinfo {author} {\bibfnamefont
  {D.}~\bibnamefont {Kim}}, \bibinfo {author} {\bibfnamefont {Y.}~\bibnamefont
  {Lee}}, \bibinfo {author} {\bibfnamefont {B.}~\bibnamefont {Seok}}, \bibinfo
  {author} {\bibfnamefont {J.}~\bibnamefont {Kim}}, \bibinfo {author}
  {\bibfnamefont {B.~G.}\ \bibnamefont {Park}}, \bibinfo {author}
  {\bibfnamefont {L.}~\bibnamefont {\ifmmode~\check{S}\else \v{S}\fi{}mejkal}},
  \bibinfo {author} {\bibfnamefont {C.-J.}\ \bibnamefont {Kang}},\ and\
  \bibinfo {author} {\bibfnamefont {C.}~\bibnamefont {Kim}},\ }\bibfield
  {title} {\bibinfo {title} {{Broken Kramers Degeneracy in Altermagnetic
  MnTe}},\ }\href {https://doi.org/10.1103/PhysRevLett.132.036702} {\bibfield
  {journal} {\bibinfo  {journal} {Phys. Rev. Lett.}\ }\textbf {\bibinfo
  {volume} {132}},\ \bibinfo {pages} {036702} (\bibinfo {year}
  {2024})}\BibitemShut {NoStop}%
\bibitem [{\citenamefont {Krempask{\'y}}\ \emph {et~al.}(2024)\citenamefont
  {Krempask{\'y}}, \citenamefont {{\v{S}}mejkal}, \citenamefont {D'Souza},
  \citenamefont {Hajlaoui}, \citenamefont {Springholz}, \citenamefont
  {Uhl{\'i}{\v{r}}ov{\'a}}, \citenamefont {Alarab}, \citenamefont
  {Constantinou}, \citenamefont {Strocov}, \citenamefont {Usanov},
  \citenamefont {Pudelko}, \citenamefont {Gonz{\'a}lez-Hern{\'a}ndez},
  \citenamefont {Birk~Hellenes}, \citenamefont {Jansa}, \citenamefont
  {Reichlov{\'a}}, \citenamefont {{\v{S}}ob{\'a}{\v{n}}}, \citenamefont
  {Gonzalez~Betancourt}, \citenamefont {Wadley}, \citenamefont {Sinova},
  \citenamefont {Kriegner}, \citenamefont {Min{\'a}r}, \citenamefont {Dil},\
  and\ \citenamefont {Jungwirth}}]{Krempasky2024}%
  \BibitemOpen
  \bibfield  {author} {\bibinfo {author} {\bibfnamefont {J.}~\bibnamefont
  {Krempask{\'y}}}, \bibinfo {author} {\bibfnamefont {L.}~\bibnamefont
  {{\v{S}}mejkal}}, \bibinfo {author} {\bibfnamefont {S.~W.}\ \bibnamefont
  {D'Souza}}, \bibinfo {author} {\bibfnamefont {M.}~\bibnamefont {Hajlaoui}},
  \bibinfo {author} {\bibfnamefont {G.}~\bibnamefont {Springholz}}, \bibinfo
  {author} {\bibfnamefont {K.}~\bibnamefont {Uhl{\'i}{\v{r}}ov{\'a}}}, \bibinfo
  {author} {\bibfnamefont {F.}~\bibnamefont {Alarab}}, \bibinfo {author}
  {\bibfnamefont {P.~C.}\ \bibnamefont {Constantinou}}, \bibinfo {author}
  {\bibfnamefont {V.}~\bibnamefont {Strocov}}, \bibinfo {author} {\bibfnamefont
  {D.}~\bibnamefont {Usanov}}, \bibinfo {author} {\bibfnamefont {W.~R.}\
  \bibnamefont {Pudelko}}, \bibinfo {author} {\bibfnamefont {R.}~\bibnamefont
  {Gonz{\'a}lez-Hern{\'a}ndez}}, \bibinfo {author} {\bibfnamefont
  {A.}~\bibnamefont {Birk~Hellenes}}, \bibinfo {author} {\bibfnamefont
  {Z.}~\bibnamefont {Jansa}}, \bibinfo {author} {\bibfnamefont
  {H.}~\bibnamefont {Reichlov{\'a}}}, \bibinfo {author} {\bibfnamefont
  {Z.}~\bibnamefont {{\v{S}}ob{\'a}{\v{n}}}}, \bibinfo {author} {\bibfnamefont
  {R.~D.}\ \bibnamefont {Gonzalez~Betancourt}}, \bibinfo {author}
  {\bibfnamefont {P.}~\bibnamefont {Wadley}}, \bibinfo {author} {\bibfnamefont
  {J.}~\bibnamefont {Sinova}}, \bibinfo {author} {\bibfnamefont
  {D.}~\bibnamefont {Kriegner}}, \bibinfo {author} {\bibfnamefont
  {J.}~\bibnamefont {Min{\'a}r}}, \bibinfo {author} {\bibfnamefont {J.~H.}\
  \bibnamefont {Dil}},\ and\ \bibinfo {author} {\bibfnamefont {T.}~\bibnamefont
  {Jungwirth}},\ }\bibfield  {title} {\bibinfo {title} {{Altermagnetic lifting
  of Kramers spin degeneracy}},\ }\href
  {https://doi.org/10.1038/s41586-023-06907-7} {\bibfield  {journal} {\bibinfo
  {journal} {Nature}\ }\textbf {\bibinfo {volume} {626}},\ \bibinfo {pages}
  {517} (\bibinfo {year} {2024})}\BibitemShut {NoStop}%
\bibitem [{\citenamefont {Hajlaoui}\ \emph {et~al.}(2024)\citenamefont
  {Hajlaoui}, \citenamefont {Wilfred~D'Souza}, \citenamefont {Šmejkal},
  \citenamefont {Kriegner}, \citenamefont {Krizman}, \citenamefont {Zakusylo},
  \citenamefont {Olszowska}, \citenamefont {Caha}, \citenamefont {Michalička},
  \citenamefont {Sánchez-Barriga}, \citenamefont {Marmodoro}, \citenamefont
  {Výborný}, \citenamefont {Ernst}, \citenamefont {Cinchetti}, \citenamefont
  {Minar}, \citenamefont {Jungwirth},\ and\ \citenamefont
  {Springholz}}]{Hajlaoui2024AM}%
  \BibitemOpen
  \bibfield  {author} {\bibinfo {author} {\bibfnamefont {M.}~\bibnamefont
  {Hajlaoui}}, \bibinfo {author} {\bibfnamefont {S.}~\bibnamefont
  {Wilfred~D'Souza}}, \bibinfo {author} {\bibfnamefont {L.}~\bibnamefont
  {Šmejkal}}, \bibinfo {author} {\bibfnamefont {D.}~\bibnamefont {Kriegner}},
  \bibinfo {author} {\bibfnamefont {G.}~\bibnamefont {Krizman}}, \bibinfo
  {author} {\bibfnamefont {T.}~\bibnamefont {Zakusylo}}, \bibinfo {author}
  {\bibfnamefont {N.}~\bibnamefont {Olszowska}}, \bibinfo {author}
  {\bibfnamefont {O.}~\bibnamefont {Caha}}, \bibinfo {author} {\bibfnamefont
  {J.}~\bibnamefont {Michalička}}, \bibinfo {author} {\bibfnamefont
  {J.}~\bibnamefont {Sánchez-Barriga}}, \bibinfo {author} {\bibfnamefont
  {A.}~\bibnamefont {Marmodoro}}, \bibinfo {author} {\bibfnamefont
  {K.}~\bibnamefont {Výborný}}, \bibinfo {author} {\bibfnamefont
  {A.}~\bibnamefont {Ernst}}, \bibinfo {author} {\bibfnamefont
  {M.}~\bibnamefont {Cinchetti}}, \bibinfo {author} {\bibfnamefont
  {J.}~\bibnamefont {Minar}}, \bibinfo {author} {\bibfnamefont
  {T.}~\bibnamefont {Jungwirth}},\ and\ \bibinfo {author} {\bibfnamefont
  {G.}~\bibnamefont {Springholz}},\ }\bibfield  {title} {\bibinfo {title}
  {Temperature dependence of relativistic valence band splitting induced by an
  altermagnetic phase transition},\ }\href
  {https://doi.org/https://doi.org/10.1002/adma.202314076} {\bibfield
  {journal} {\bibinfo  {journal} {Advanced Materials}\ }\textbf {\bibinfo
  {volume} {36}},\ \bibinfo {pages} {2314076} (\bibinfo {year}
  {2024})}\BibitemShut {NoStop}%
\bibitem [{\citenamefont {Jiang}\ \emph {et~al.}(2025)\citenamefont {Jiang},
  \citenamefont {Hu}, \citenamefont {Bai}, \citenamefont {Song}, \citenamefont
  {Mu}, \citenamefont {Qu}, \citenamefont {Li}, \citenamefont {Zhu},
  \citenamefont {Pi}, \citenamefont {Wei}, \citenamefont {Sun}, \citenamefont
  {Huang}, \citenamefont {Zheng}, \citenamefont {Peng}, \citenamefont {He},
  \citenamefont {Li}, \citenamefont {Luo}, \citenamefont {Li}, \citenamefont
  {Chen}, \citenamefont {Li}, \citenamefont {Weng},\ and\ \citenamefont
  {Qian}}]{Jiang2024KV2Se2O}%
  \BibitemOpen
  \bibfield  {author} {\bibinfo {author} {\bibfnamefont {B.}~\bibnamefont
  {Jiang}}, \bibinfo {author} {\bibfnamefont {M.}~\bibnamefont {Hu}}, \bibinfo
  {author} {\bibfnamefont {J.}~\bibnamefont {Bai}}, \bibinfo {author}
  {\bibfnamefont {Z.}~\bibnamefont {Song}}, \bibinfo {author} {\bibfnamefont
  {C.}~\bibnamefont {Mu}}, \bibinfo {author} {\bibfnamefont {G.}~\bibnamefont
  {Qu}}, \bibinfo {author} {\bibfnamefont {W.}~\bibnamefont {Li}}, \bibinfo
  {author} {\bibfnamefont {W.}~\bibnamefont {Zhu}}, \bibinfo {author}
  {\bibfnamefont {H.}~\bibnamefont {Pi}}, \bibinfo {author} {\bibfnamefont
  {Z.}~\bibnamefont {Wei}}, \bibinfo {author} {\bibfnamefont {Y.-J.}\
  \bibnamefont {Sun}}, \bibinfo {author} {\bibfnamefont {Y.}~\bibnamefont
  {Huang}}, \bibinfo {author} {\bibfnamefont {X.}~\bibnamefont {Zheng}},
  \bibinfo {author} {\bibfnamefont {Y.}~\bibnamefont {Peng}}, \bibinfo {author}
  {\bibfnamefont {L.}~\bibnamefont {He}}, \bibinfo {author} {\bibfnamefont
  {S.}~\bibnamefont {Li}}, \bibinfo {author} {\bibfnamefont {J.}~\bibnamefont
  {Luo}}, \bibinfo {author} {\bibfnamefont {Z.}~\bibnamefont {Li}}, \bibinfo
  {author} {\bibfnamefont {G.}~\bibnamefont {Chen}}, \bibinfo {author}
  {\bibfnamefont {H.}~\bibnamefont {Li}}, \bibinfo {author} {\bibfnamefont
  {H.}~\bibnamefont {Weng}},\ and\ \bibinfo {author} {\bibfnamefont
  {T.}~\bibnamefont {Qian}},\ }\bibfield  {title} {\bibinfo {title} {{A
  metallic room-temperature d-wave altermagnet}},\ }\href
  {https://doi.org/10.1038/s41567-025-02822-y} {\bibfield  {journal} {\bibinfo
  {journal} {Nature Physics}\ }\textbf {\bibinfo {volume} {21}},\ \bibinfo
  {pages} {754} (\bibinfo {year} {2025})}\BibitemShut {NoStop}%
\bibitem [{\citenamefont {Zhang}\ \emph {et~al.}(2025)\citenamefont {Zhang},
  \citenamefont {Cheng}, \citenamefont {Yin}, \citenamefont {Liu},
  \citenamefont {Deng}, \citenamefont {Qiao}, \citenamefont {Shi},
  \citenamefont {Zhang}, \citenamefont {Lin}, \citenamefont {Liu},
  \citenamefont {Ye}, \citenamefont {Huang}, \citenamefont {Meng},
  \citenamefont {Zhang}, \citenamefont {Okuda}, \citenamefont {Shimada},
  \citenamefont {Cui}, \citenamefont {Zhao}, \citenamefont {Cao}, \citenamefont
  {Qiao}, \citenamefont {Liu},\ and\ \citenamefont {Chen}}]{Zhang2025Cpair}%
  \BibitemOpen
  \bibfield  {author} {\bibinfo {author} {\bibfnamefont {F.}~\bibnamefont
  {Zhang}}, \bibinfo {author} {\bibfnamefont {X.}~\bibnamefont {Cheng}},
  \bibinfo {author} {\bibfnamefont {Z.}~\bibnamefont {Yin}}, \bibinfo {author}
  {\bibfnamefont {C.}~\bibnamefont {Liu}}, \bibinfo {author} {\bibfnamefont
  {L.}~\bibnamefont {Deng}}, \bibinfo {author} {\bibfnamefont {Y.}~\bibnamefont
  {Qiao}}, \bibinfo {author} {\bibfnamefont {Z.}~\bibnamefont {Shi}}, \bibinfo
  {author} {\bibfnamefont {S.}~\bibnamefont {Zhang}}, \bibinfo {author}
  {\bibfnamefont {J.}~\bibnamefont {Lin}}, \bibinfo {author} {\bibfnamefont
  {Z.}~\bibnamefont {Liu}}, \bibinfo {author} {\bibfnamefont {M.}~\bibnamefont
  {Ye}}, \bibinfo {author} {\bibfnamefont {Y.}~\bibnamefont {Huang}}, \bibinfo
  {author} {\bibfnamefont {X.}~\bibnamefont {Meng}}, \bibinfo {author}
  {\bibfnamefont {C.}~\bibnamefont {Zhang}}, \bibinfo {author} {\bibfnamefont
  {T.}~\bibnamefont {Okuda}}, \bibinfo {author} {\bibfnamefont
  {K.}~\bibnamefont {Shimada}}, \bibinfo {author} {\bibfnamefont
  {S.}~\bibnamefont {Cui}}, \bibinfo {author} {\bibfnamefont {Y.}~\bibnamefont
  {Zhao}}, \bibinfo {author} {\bibfnamefont {G.~H.}\ \bibnamefont {Cao}},
  \bibinfo {author} {\bibfnamefont {S.}~\bibnamefont {Qiao}}, \bibinfo {author}
  {\bibfnamefont {J.}~\bibnamefont {Liu}},\ and\ \bibinfo {author}
  {\bibfnamefont {C.}~\bibnamefont {Chen}},\ }\bibfield  {title} {\bibinfo
  {title} {Crystal-symmetry-paired spin–valley locking in a layered
  room-temperature metallic altermagnet candidate},\ }\href
  {https://doi.org/10.1038/s41567-025-02864-2} {\bibfield  {journal} {\bibinfo
  {journal} {Nature Physics}\ }\textbf {\bibinfo {volume} {21}},\ \bibinfo
  {pages} {760} (\bibinfo {year} {2025})}\BibitemShut {NoStop}%
\bibitem [{\citenamefont {Wang}\ \emph {et~al.}(2025)\citenamefont {Wang},
  \citenamefont {Yu}, \citenamefont {Cheng}, \citenamefont {Xiao},
  \citenamefont {Ma}, \citenamefont {Quan}, \citenamefont {Song}, \citenamefont
  {Zhang}, \citenamefont {Zhang}, \citenamefont {Ma}, \citenamefont {Liu},
  \citenamefont {Yadav}, \citenamefont {Shi}, \citenamefont {Wang},
  \citenamefont {Niu}, \citenamefont {Gao}, \citenamefont {Xiang},
  \citenamefont {Liu}, \citenamefont {Wang},\ and\ \citenamefont
  {Chen}}]{wang2025KVSeO}%
  \BibitemOpen
  \bibfield  {author} {\bibinfo {author} {\bibfnamefont {Z.}~\bibnamefont
  {Wang}}, \bibinfo {author} {\bibfnamefont {S.}~\bibnamefont {Yu}}, \bibinfo
  {author} {\bibfnamefont {X.}~\bibnamefont {Cheng}}, \bibinfo {author}
  {\bibfnamefont {X.}~\bibnamefont {Xiao}}, \bibinfo {author} {\bibfnamefont
  {W.}~\bibnamefont {Ma}}, \bibinfo {author} {\bibfnamefont {F.}~\bibnamefont
  {Quan}}, \bibinfo {author} {\bibfnamefont {H.}~\bibnamefont {Song}}, \bibinfo
  {author} {\bibfnamefont {K.}~\bibnamefont {Zhang}}, \bibinfo {author}
  {\bibfnamefont {Y.}~\bibnamefont {Zhang}}, \bibinfo {author} {\bibfnamefont
  {Y.}~\bibnamefont {Ma}}, \bibinfo {author} {\bibfnamefont {W.}~\bibnamefont
  {Liu}}, \bibinfo {author} {\bibfnamefont {P.}~\bibnamefont {Yadav}}, \bibinfo
  {author} {\bibfnamefont {X.}~\bibnamefont {Shi}}, \bibinfo {author}
  {\bibfnamefont {Z.}~\bibnamefont {Wang}}, \bibinfo {author} {\bibfnamefont
  {Q.}~\bibnamefont {Niu}}, \bibinfo {author} {\bibfnamefont {Y.}~\bibnamefont
  {Gao}}, \bibinfo {author} {\bibfnamefont {B.}~\bibnamefont {Xiang}}, \bibinfo
  {author} {\bibfnamefont {J.}~\bibnamefont {Liu}}, \bibinfo {author}
  {\bibfnamefont {Z.}~\bibnamefont {Wang}},\ and\ \bibinfo {author}
  {\bibfnamefont {X.}~\bibnamefont {Chen}},\ }\href
  {https://arxiv.org/abs/2512.23290} {\bibinfo {title} {Atomic-scale spin
  sensing of a 2d $d$-wave altermagnet via helical tunneling}} (\bibinfo {year}
  {2025}),\ \Eprint {https://arxiv.org/abs/2512.23290} {arXiv:2512.23290
  [cond-mat.mes-hall]} \BibitemShut {NoStop}%
\bibitem [{\citenamefont {Yang}\ \emph {et~al.}(2026)\citenamefont {Yang},
  \citenamefont {Li}, \citenamefont {Wang}, \citenamefont {Zhao}, \citenamefont
  {Wan}, \citenamefont {Gui}, \citenamefont {Zeng}, \citenamefont {Cao},
  \citenamefont {Hu}, \citenamefont {Yu}, \citenamefont {Zhang}, \citenamefont
  {Chen}, \citenamefont {Liu}, \citenamefont {Song}, \citenamefont {Zhang},
  \citenamefont {Liu}, \citenamefont {Hu}, \citenamefont {Jiao},\ and\
  \citenamefont {Yuan}}]{Jiao2026KVSeO}%
  \BibitemOpen
  \bibfield  {author} {\bibinfo {author} {\bibfnamefont {G.}~\bibnamefont
  {Yang}}, \bibinfo {author} {\bibfnamefont {C.}~\bibnamefont {Li}}, \bibinfo
  {author} {\bibfnamefont {C.}~\bibnamefont {Wang}}, \bibinfo {author}
  {\bibfnamefont {X.}~\bibnamefont {Zhao}}, \bibinfo {author} {\bibfnamefont
  {Y.}~\bibnamefont {Wan}}, \bibinfo {author} {\bibfnamefont {H.}~\bibnamefont
  {Gui}}, \bibinfo {author} {\bibfnamefont {G.}~\bibnamefont {Zeng}}, \bibinfo
  {author} {\bibfnamefont {S.}~\bibnamefont {Cao}}, \bibinfo {author}
  {\bibfnamefont {C.}~\bibnamefont {Hu}}, \bibinfo {author} {\bibfnamefont
  {Q.}~\bibnamefont {Yu}}, \bibinfo {author} {\bibfnamefont {Y.}~\bibnamefont
  {Zhang}}, \bibinfo {author} {\bibfnamefont {D.}~\bibnamefont {Chen}},
  \bibinfo {author} {\bibfnamefont {Y.}~\bibnamefont {Liu}}, \bibinfo {author}
  {\bibfnamefont {Y.}~\bibnamefont {Song}}, \bibinfo {author} {\bibfnamefont
  {Y.}~\bibnamefont {Zhang}}, \bibinfo {author} {\bibfnamefont
  {F.}~\bibnamefont {Liu}}, \bibinfo {author} {\bibfnamefont {L.-H.}\
  \bibnamefont {Hu}}, \bibinfo {author} {\bibfnamefont {L.}~\bibnamefont
  {Jiao}},\ and\ \bibinfo {author} {\bibfnamefont {H.}~\bibnamefont {Yuan}},\
  }\href {https://arxiv.org/abs/2603.21969} {\bibinfo {title} {Visualizing
  spin-polarization of an altermagnet kv$_2$se$_2$o via spin-selective
  tunneling}} (\bibinfo {year} {2026}),\ \Eprint
  {https://arxiv.org/abs/2603.21969} {arXiv:2603.21969 [cond-mat.mtrl-sci]}
  \BibitemShut {NoStop}%
\bibitem [{\citenamefont {\ifmmode~\check{S}\else \v{S}\fi{}mejkal}\ \emph
  {et~al.}(2022{\natexlab{b}})\citenamefont {\ifmmode~\check{S}\else
  \v{S}\fi{}mejkal}, \citenamefont {Sinova},\ and\ \citenamefont
  {Jungwirth}}]{Libor2022AMb}%
  \BibitemOpen
  \bibfield  {author} {\bibinfo {author} {\bibfnamefont {L.}~\bibnamefont
  {\ifmmode~\check{S}\else \v{S}\fi{}mejkal}}, \bibinfo {author} {\bibfnamefont
  {J.}~\bibnamefont {Sinova}},\ and\ \bibinfo {author} {\bibfnamefont
  {T.}~\bibnamefont {Jungwirth}},\ }\bibfield  {title} {\bibinfo {title}
  {{Emerging Research Landscape of Altermagnetism}},\ }\href
  {https://doi.org/10.1103/PhysRevX.12.040501} {\bibfield  {journal} {\bibinfo
  {journal} {Phys. Rev. X}\ }\textbf {\bibinfo {volume} {12}},\ \bibinfo
  {pages} {040501} (\bibinfo {year} {2022}{\natexlab{b}})}\BibitemShut
  {NoStop}%
\bibitem [{\citenamefont {Chen}\ \emph {et~al.}(2026)\citenamefont {Chen},
  \citenamefont {Chen},\ and\ \citenamefont {Liu}}]{chen2026review}%
  \BibitemOpen
  \bibfield  {author} {\bibinfo {author} {\bibfnamefont {X.}~\bibnamefont
  {Chen}}, \bibinfo {author} {\bibfnamefont {W.}~\bibnamefont {Chen}},\ and\
  \bibinfo {author} {\bibfnamefont {Q.}~\bibnamefont {Liu}},\ }\href
  {https://arxiv.org/abs/2603.27505} {\bibinfo {title} {The rise of
  unconventional magnetism}} (\bibinfo {year} {2026}),\ \Eprint
  {https://arxiv.org/abs/2603.27505} {arXiv:2603.27505 [cond-mat.mtrl-sci]}
  \BibitemShut {NoStop}%
\bibitem [{\citenamefont {Bai}\ \emph {et~al.}(2023)\citenamefont {Bai},
  \citenamefont {Zhang}, \citenamefont {Zhou}, \citenamefont {Chen},
  \citenamefont {Wan}, \citenamefont {Han}, \citenamefont {Zhu}, \citenamefont
  {Liang}, \citenamefont {Su}, \citenamefont {Han}, \citenamefont {Pan},\ and\
  \citenamefont {Song}}]{Bai2023AM}%
  \BibitemOpen
  \bibfield  {author} {\bibinfo {author} {\bibfnamefont {H.}~\bibnamefont
  {Bai}}, \bibinfo {author} {\bibfnamefont {Y.~C.}\ \bibnamefont {Zhang}},
  \bibinfo {author} {\bibfnamefont {Y.~J.}\ \bibnamefont {Zhou}}, \bibinfo
  {author} {\bibfnamefont {P.}~\bibnamefont {Chen}}, \bibinfo {author}
  {\bibfnamefont {C.~H.}\ \bibnamefont {Wan}}, \bibinfo {author} {\bibfnamefont
  {L.}~\bibnamefont {Han}}, \bibinfo {author} {\bibfnamefont {W.~X.}\
  \bibnamefont {Zhu}}, \bibinfo {author} {\bibfnamefont {S.~X.}\ \bibnamefont
  {Liang}}, \bibinfo {author} {\bibfnamefont {Y.~C.}\ \bibnamefont {Su}},
  \bibinfo {author} {\bibfnamefont {X.~F.}\ \bibnamefont {Han}}, \bibinfo
  {author} {\bibfnamefont {F.}~\bibnamefont {Pan}},\ and\ \bibinfo {author}
  {\bibfnamefont {C.}~\bibnamefont {Song}},\ }\bibfield  {title} {\bibinfo
  {title} {Efficient spin-to-charge conversion via altermagnetic spin splitting
  effect in antiferromagnet ${\mathrm{ruo}}_{2}$},\ }\href
  {https://doi.org/10.1103/PhysRevLett.130.216701} {\bibfield  {journal}
  {\bibinfo  {journal} {Phys. Rev. Lett.}\ }\textbf {\bibinfo {volume} {130}},\
  \bibinfo {pages} {216701} (\bibinfo {year} {2023})}\BibitemShut {NoStop}%
\bibitem [{\citenamefont {Han}\ \emph {et~al.}(2024)\citenamefont {Han},
  \citenamefont {Fu}, \citenamefont {Peng}, \citenamefont {Cheng},
  \citenamefont {Dai}, \citenamefont {Liu}, \citenamefont {Li}, \citenamefont
  {Zhang}, \citenamefont {Zhu}, \citenamefont {Bai}, \citenamefont {Zhou},
  \citenamefont {Liang}, \citenamefont {Chen}, \citenamefont {Wang},
  \citenamefont {Chen}, \citenamefont {Yang}, \citenamefont {Zhang},
  \citenamefont {Song}, \citenamefont {Liu},\ and\ \citenamefont
  {Pan}}]{Han2024AM}%
  \BibitemOpen
  \bibfield  {author} {\bibinfo {author} {\bibfnamefont {L.}~\bibnamefont
  {Han}}, \bibinfo {author} {\bibfnamefont {X.}~\bibnamefont {Fu}}, \bibinfo
  {author} {\bibfnamefont {R.}~\bibnamefont {Peng}}, \bibinfo {author}
  {\bibfnamefont {X.}~\bibnamefont {Cheng}}, \bibinfo {author} {\bibfnamefont
  {J.}~\bibnamefont {Dai}}, \bibinfo {author} {\bibfnamefont {L.}~\bibnamefont
  {Liu}}, \bibinfo {author} {\bibfnamefont {Y.}~\bibnamefont {Li}}, \bibinfo
  {author} {\bibfnamefont {Y.}~\bibnamefont {Zhang}}, \bibinfo {author}
  {\bibfnamefont {W.}~\bibnamefont {Zhu}}, \bibinfo {author} {\bibfnamefont
  {H.}~\bibnamefont {Bai}}, \bibinfo {author} {\bibfnamefont {Y.}~\bibnamefont
  {Zhou}}, \bibinfo {author} {\bibfnamefont {S.}~\bibnamefont {Liang}},
  \bibinfo {author} {\bibfnamefont {C.}~\bibnamefont {Chen}}, \bibinfo {author}
  {\bibfnamefont {Q.}~\bibnamefont {Wang}}, \bibinfo {author} {\bibfnamefont
  {X.}~\bibnamefont {Chen}}, \bibinfo {author} {\bibfnamefont {L.}~\bibnamefont
  {Yang}}, \bibinfo {author} {\bibfnamefont {Y.}~\bibnamefont {Zhang}},
  \bibinfo {author} {\bibfnamefont {C.}~\bibnamefont {Song}}, \bibinfo {author}
  {\bibfnamefont {J.}~\bibnamefont {Liu}},\ and\ \bibinfo {author}
  {\bibfnamefont {F.}~\bibnamefont {Pan}},\ }\bibfield  {title} {\bibinfo
  {title} {{Electrical 180$^{\circ}$ switching of N\'{e}el vector in
  spin-splitting antiferromagnet}},\ }\href
  {https://doi.org/10.1126/sciadv.adn0479} {\bibfield  {journal} {\bibinfo
  {journal} {Science Advances}\ }\textbf {\bibinfo {volume} {10}},\ \bibinfo
  {pages} {eadn0479} (\bibinfo {year} {2024})}\BibitemShut {NoStop}%
\bibitem [{\citenamefont {Hu}\ \emph {et~al.}(2025)\citenamefont {Hu},
  \citenamefont {Matsyshyn},\ and\ \citenamefont {Song}}]{Hu2025NLME}%
  \BibitemOpen
  \bibfield  {author} {\bibinfo {author} {\bibfnamefont {J.-X.}\ \bibnamefont
  {Hu}}, \bibinfo {author} {\bibfnamefont {O.}~\bibnamefont {Matsyshyn}},\ and\
  \bibinfo {author} {\bibfnamefont {J.~C.~W.}\ \bibnamefont {Song}},\
  }\bibfield  {title} {\bibinfo {title} {{Nonlinear Superconducting
  Magnetoelectric Effect}},\ }\href
  {https://doi.org/10.1103/PhysRevLett.134.026001} {\bibfield  {journal}
  {\bibinfo  {journal} {Phys. Rev. Lett.}\ }\textbf {\bibinfo {volume} {134}},\
  \bibinfo {pages} {026001} (\bibinfo {year} {2025})}\BibitemShut {NoStop}%
\bibitem [{\citenamefont {Duan}\ \emph {et~al.}(2025)\citenamefont {Duan},
  \citenamefont {Zhang}, \citenamefont {Zhu}, \citenamefont {Liu},
  \citenamefont {Zhang}, \citenamefont {\ifmmode \check{Z}\else
  \v{Z}\fi{}uti\ifmmode~\acute{c}\else \'{c}\fi{}},\ and\ \citenamefont
  {Zhou}}]{Duan2025AFMAM}%
  \BibitemOpen
  \bibfield  {author} {\bibinfo {author} {\bibfnamefont {X.}~\bibnamefont
  {Duan}}, \bibinfo {author} {\bibfnamefont {J.}~\bibnamefont {Zhang}},
  \bibinfo {author} {\bibfnamefont {Z.}~\bibnamefont {Zhu}}, \bibinfo {author}
  {\bibfnamefont {Y.}~\bibnamefont {Liu}}, \bibinfo {author} {\bibfnamefont
  {Z.}~\bibnamefont {Zhang}}, \bibinfo {author} {\bibfnamefont
  {I.}~\bibnamefont {\ifmmode \check{Z}\else
  \v{Z}\fi{}uti\ifmmode~\acute{c}\else \'{c}\fi{}}},\ and\ \bibinfo {author}
  {\bibfnamefont {T.}~\bibnamefont {Zhou}},\ }\bibfield  {title} {\bibinfo
  {title} {{Antiferroelectric Altermagnets: Antiferroelectricity Alters
  Magnets}},\ }\href {https://doi.org/10.1103/PhysRevLett.134.106801}
  {\bibfield  {journal} {\bibinfo  {journal} {Phys. Rev. Lett.}\ }\textbf
  {\bibinfo {volume} {134}},\ \bibinfo {pages} {106801} (\bibinfo {year}
  {2025})}\BibitemShut {NoStop}%
\bibitem [{\citenamefont {Gu}\ \emph {et~al.}(2025)\citenamefont {Gu},
  \citenamefont {Liu}, \citenamefont {Zhu}, \citenamefont {Yananose},
  \citenamefont {Chen}, \citenamefont {Hu}, \citenamefont {Stroppa},\ and\
  \citenamefont {Liu}}]{Gu2025FEAM}%
  \BibitemOpen
  \bibfield  {author} {\bibinfo {author} {\bibfnamefont {M.}~\bibnamefont
  {Gu}}, \bibinfo {author} {\bibfnamefont {Y.}~\bibnamefont {Liu}}, \bibinfo
  {author} {\bibfnamefont {H.}~\bibnamefont {Zhu}}, \bibinfo {author}
  {\bibfnamefont {K.}~\bibnamefont {Yananose}}, \bibinfo {author}
  {\bibfnamefont {X.}~\bibnamefont {Chen}}, \bibinfo {author} {\bibfnamefont
  {Y.}~\bibnamefont {Hu}}, \bibinfo {author} {\bibfnamefont {A.}~\bibnamefont
  {Stroppa}},\ and\ \bibinfo {author} {\bibfnamefont {Q.}~\bibnamefont {Liu}},\
  }\bibfield  {title} {\bibinfo {title} {{Ferroelectric Switchable
  Altermagnetism}},\ }\href {https://doi.org/10.1103/PhysRevLett.134.106802}
  {\bibfield  {journal} {\bibinfo  {journal} {Phys. Rev. Lett.}\ }\textbf
  {\bibinfo {volume} {134}},\ \bibinfo {pages} {106802} (\bibinfo {year}
  {2025})}\BibitemShut {NoStop}%
\bibitem [{\citenamefont {Jungwirth}\ \emph {et~al.}(2025)\citenamefont
  {Jungwirth}, \citenamefont {Sinova}, \citenamefont {Wadley}, \citenamefont
  {Kriegner}, \citenamefont {Reichlova}, \citenamefont {Krizek}, \citenamefont
  {Ohno},\ and\ \citenamefont {Smejkal}}]{Jungwirth2025review}%
  \BibitemOpen
  \bibfield  {author} {\bibinfo {author} {\bibfnamefont {T.}~\bibnamefont
  {Jungwirth}}, \bibinfo {author} {\bibfnamefont {J.}~\bibnamefont {Sinova}},
  \bibinfo {author} {\bibfnamefont {P.}~\bibnamefont {Wadley}}, \bibinfo
  {author} {\bibfnamefont {D.}~\bibnamefont {Kriegner}}, \bibinfo {author}
  {\bibfnamefont {H.}~\bibnamefont {Reichlova}}, \bibinfo {author}
  {\bibfnamefont {F.}~\bibnamefont {Krizek}}, \bibinfo {author} {\bibfnamefont
  {H.}~\bibnamefont {Ohno}},\ and\ \bibinfo {author} {\bibfnamefont
  {L.}~\bibnamefont {Smejkal}},\ }\href {https://arxiv.org/abs/2508.09748}
  {\bibinfo {title} {Altermagnetic spintronics}} (\bibinfo {year} {2025}),\
  \Eprint {https://arxiv.org/abs/2508.09748} {arXiv:2508.09748
  [cond-mat.mtrl-sci]} \BibitemShut {NoStop}%
\bibitem [{\citenamefont {Zhu}\ \emph {et~al.}(2023)\citenamefont {Zhu},
  \citenamefont {Zhuang}, \citenamefont {Wu},\ and\ \citenamefont
  {Yan}}]{Zhu2023TSC}%
  \BibitemOpen
  \bibfield  {author} {\bibinfo {author} {\bibfnamefont {D.}~\bibnamefont
  {Zhu}}, \bibinfo {author} {\bibfnamefont {Z.-Y.}\ \bibnamefont {Zhuang}},
  \bibinfo {author} {\bibfnamefont {Z.}~\bibnamefont {Wu}},\ and\ \bibinfo
  {author} {\bibfnamefont {Z.}~\bibnamefont {Yan}},\ }\bibfield  {title}
  {\bibinfo {title} {Topological superconductivity in two-dimensional
  altermagnetic metals},\ }\href {https://doi.org/10.1103/PhysRevB.108.184505}
  {\bibfield  {journal} {\bibinfo  {journal} {Phys. Rev. B}\ }\textbf {\bibinfo
  {volume} {108}},\ \bibinfo {pages} {184505} (\bibinfo {year}
  {2023})}\BibitemShut {NoStop}%
\bibitem [{\citenamefont {Zhu}\ \emph {et~al.}(2024)\citenamefont {Zhu},
  \citenamefont {Liu}, \citenamefont {Zhuang}, \citenamefont {Wu},\ and\
  \citenamefont {Yan}}]{Zhu2024dislocation}%
  \BibitemOpen
  \bibfield  {author} {\bibinfo {author} {\bibfnamefont {D.}~\bibnamefont
  {Zhu}}, \bibinfo {author} {\bibfnamefont {D.}~\bibnamefont {Liu}}, \bibinfo
  {author} {\bibfnamefont {Z.-Y.}\ \bibnamefont {Zhuang}}, \bibinfo {author}
  {\bibfnamefont {Z.}~\bibnamefont {Wu}},\ and\ \bibinfo {author}
  {\bibfnamefont {Z.}~\bibnamefont {Yan}},\ }\bibfield  {title} {\bibinfo
  {title} {{Field-sensitive dislocation bound states in two-dimensional
  $d$-wave altermagnets}},\ }\href
  {https://doi.org/10.1103/PhysRevB.110.165141} {\bibfield  {journal} {\bibinfo
   {journal} {Phys. Rev. B}\ }\textbf {\bibinfo {volume} {110}},\ \bibinfo
  {pages} {165141} (\bibinfo {year} {2024})}\BibitemShut {NoStop}%
\bibitem [{\citenamefont {Li}\ and\ \citenamefont {Liu}(2023)}]{Li2023AMHOTSC}%
  \BibitemOpen
  \bibfield  {author} {\bibinfo {author} {\bibfnamefont {Y.-X.}\ \bibnamefont
  {Li}}\ and\ \bibinfo {author} {\bibfnamefont {C.-C.}\ \bibnamefont {Liu}},\
  }\bibfield  {title} {\bibinfo {title} {Majorana corner modes and tunable
  patterns in an altermagnet heterostructure},\ }\href
  {https://doi.org/10.1103/PhysRevB.108.205410} {\bibfield  {journal} {\bibinfo
   {journal} {Phys. Rev. B}\ }\textbf {\bibinfo {volume} {108}},\ \bibinfo
  {pages} {205410} (\bibinfo {year} {2023})}\BibitemShut {NoStop}%
\bibitem [{\citenamefont {Li}\ \emph {et~al.}(2024)\citenamefont {Li},
  \citenamefont {Liu},\ and\ \citenamefont {Liu}}]{Li2024AMHOTI}%
  \BibitemOpen
  \bibfield  {author} {\bibinfo {author} {\bibfnamefont {Y.-X.}\ \bibnamefont
  {Li}}, \bibinfo {author} {\bibfnamefont {Y.}~\bibnamefont {Liu}},\ and\
  \bibinfo {author} {\bibfnamefont {C.-C.}\ \bibnamefont {Liu}},\ }\bibfield
  {title} {\bibinfo {title} {Creation and manipulation of higher-order
  topological states by altermagnets},\ }\href
  {https://doi.org/10.1103/PhysRevB.109.L201109} {\bibfield  {journal}
  {\bibinfo  {journal} {Phys. Rev. B}\ }\textbf {\bibinfo {volume} {109}},\
  \bibinfo {pages} {L201109} (\bibinfo {year} {2024})}\BibitemShut {NoStop}%
\bibitem [{\citenamefont {Ghorashi}\ \emph {et~al.}(2024)\citenamefont
  {Ghorashi}, \citenamefont {Hughes},\ and\ \citenamefont
  {Cano}}]{Ghorashi2024AM}%
  \BibitemOpen
  \bibfield  {author} {\bibinfo {author} {\bibfnamefont {S.~A.~A.}\
  \bibnamefont {Ghorashi}}, \bibinfo {author} {\bibfnamefont {T.~L.}\
  \bibnamefont {Hughes}},\ and\ \bibinfo {author} {\bibfnamefont
  {J.}~\bibnamefont {Cano}},\ }\bibfield  {title} {\bibinfo {title}
  {{Altermagnetic Routes to Majorana Modes in Zero Net Magnetization}},\ }\href
  {https://doi.org/10.1103/PhysRevLett.133.106601} {\bibfield  {journal}
  {\bibinfo  {journal} {Phys. Rev. Lett.}\ }\textbf {\bibinfo {volume} {133}},\
  \bibinfo {pages} {106601} (\bibinfo {year} {2024})}\BibitemShut {NoStop}%
\bibitem [{\citenamefont {Antonenko}\ \emph {et~al.}(2025)\citenamefont
  {Antonenko}, \citenamefont {Fernandes},\ and\ \citenamefont
  {Venderbos}}]{Antonenko2024AM}%
  \BibitemOpen
  \bibfield  {author} {\bibinfo {author} {\bibfnamefont {D.~S.}\ \bibnamefont
  {Antonenko}}, \bibinfo {author} {\bibfnamefont {R.~M.}\ \bibnamefont
  {Fernandes}},\ and\ \bibinfo {author} {\bibfnamefont {J.~W.~F.}\ \bibnamefont
  {Venderbos}},\ }\bibfield  {title} {\bibinfo {title} {{Mirror Chern Bands and
  Weyl Nodal Loops in Altermagnets}},\ }\href
  {https://doi.org/10.1103/PhysRevLett.134.096703} {\bibfield  {journal}
  {\bibinfo  {journal} {Phys. Rev. Lett.}\ }\textbf {\bibinfo {volume} {134}},\
  \bibinfo {pages} {096703} (\bibinfo {year} {2025})}\BibitemShut {NoStop}%
\bibitem [{\citenamefont {Zhang}\ \emph {et~al.}(2024)\citenamefont {Zhang},
  \citenamefont {Hu},\ and\ \citenamefont {Neupert}}]{Zhang2024AM}%
  \BibitemOpen
  \bibfield  {author} {\bibinfo {author} {\bibfnamefont {S.-B.}\ \bibnamefont
  {Zhang}}, \bibinfo {author} {\bibfnamefont {L.-H.}\ \bibnamefont {Hu}},\ and\
  \bibinfo {author} {\bibfnamefont {T.}~\bibnamefont {Neupert}},\ }\bibfield
  {title} {\bibinfo {title} {{Finite-momentum Cooper pairing in proximitized
  altermagnets}},\ }\href {https://doi.org/10.1038/s41467-024-45951-3}
  {\bibfield  {journal} {\bibinfo  {journal} {Nature Communications}\ }\textbf
  {\bibinfo {volume} {15}},\ \bibinfo {pages} {1801} (\bibinfo {year}
  {2024})}\BibitemShut {NoStop}%
\bibitem [{\citenamefont {Brekke}\ \emph {et~al.}(2023)\citenamefont {Brekke},
  \citenamefont {Brataas},\ and\ \citenamefont {Sudb\o{}}}]{Brekke2023AM}%
  \BibitemOpen
  \bibfield  {author} {\bibinfo {author} {\bibfnamefont {B.}~\bibnamefont
  {Brekke}}, \bibinfo {author} {\bibfnamefont {A.}~\bibnamefont {Brataas}},\
  and\ \bibinfo {author} {\bibfnamefont {A.}~\bibnamefont {Sudb\o{}}},\
  }\bibfield  {title} {\bibinfo {title} {{Two-dimensional altermagnets:
  Superconductivity in a minimal microscopic model}},\ }\href
  {https://doi.org/10.1103/PhysRevB.108.224421} {\bibfield  {journal} {\bibinfo
   {journal} {Phys. Rev. B}\ }\textbf {\bibinfo {volume} {108}},\ \bibinfo
  {pages} {224421} (\bibinfo {year} {2023})}\BibitemShut {NoStop}%
\bibitem [{\citenamefont {Monkman}\ \emph {et~al.}(2026)\citenamefont
  {Monkman}, \citenamefont {Weng}, \citenamefont {Heinsdorf}, \citenamefont
  {Nocera}, \citenamefont {Barlas},\ and\ \citenamefont {Franz}}]{Monkman2026}%
  \BibitemOpen
  \bibfield  {author} {\bibinfo {author} {\bibfnamefont {K.}~\bibnamefont
  {Monkman}}, \bibinfo {author} {\bibfnamefont {J.}~\bibnamefont {Weng}},
  \bibinfo {author} {\bibfnamefont {N.}~\bibnamefont {Heinsdorf}}, \bibinfo
  {author} {\bibfnamefont {A.}~\bibnamefont {Nocera}}, \bibinfo {author}
  {\bibfnamefont {Y.}~\bibnamefont {Barlas}},\ and\ \bibinfo {author}
  {\bibfnamefont {M.}~\bibnamefont {Franz}},\ }\bibfield  {title} {\bibinfo
  {title} {{Persistent Spin Currents in Superconducting Altermagnets}},\ }\href
  {https://doi.org/10.1103/52wh-1z5y} {\bibfield  {journal} {\bibinfo
  {journal} {Phys. Rev. X}\ }\textbf {\bibinfo {volume} {16}},\ \bibinfo
  {pages} {011057} (\bibinfo {year} {2026})}\BibitemShut {NoStop}%
\bibitem [{\citenamefont {Fukaya}\ \emph {et~al.}(2025)\citenamefont {Fukaya},
  \citenamefont {Lu}, \citenamefont {Yada}, \citenamefont {Tanaka},\ and\
  \citenamefont {Cayao}}]{Fukaya2025review}%
  \BibitemOpen
  \bibfield  {author} {\bibinfo {author} {\bibfnamefont {Y.}~\bibnamefont
  {Fukaya}}, \bibinfo {author} {\bibfnamefont {B.}~\bibnamefont {Lu}}, \bibinfo
  {author} {\bibfnamefont {K.}~\bibnamefont {Yada}}, \bibinfo {author}
  {\bibfnamefont {Y.}~\bibnamefont {Tanaka}},\ and\ \bibinfo {author}
  {\bibfnamefont {J.}~\bibnamefont {Cayao}},\ }\bibfield  {title} {\bibinfo
  {title} {Superconducting phenomena in systems with unconventional magnets},\
  }\href {https://doi.org/10.1088/1361-648X/adf1cf} {\bibfield  {journal}
  {\bibinfo  {journal} {Journal of Physics: Condensed Matter}\ }\textbf
  {\bibinfo {volume} {37}},\ \bibinfo {pages} {313003} (\bibinfo {year}
  {2025})}\BibitemShut {NoStop}%
\bibitem [{\citenamefont {Liu}\ \emph {et~al.}(2026{\natexlab{a}})\citenamefont
  {Liu}, \citenamefont {Hu},\ and\ \citenamefont {Liu}}]{Liu2025review}%
  \BibitemOpen
  \bibfield  {author} {\bibinfo {author} {\bibfnamefont {Z.}~\bibnamefont
  {Liu}}, \bibinfo {author} {\bibfnamefont {H.}~\bibnamefont {Hu}},\ and\
  \bibinfo {author} {\bibfnamefont {X.-J.}\ \bibnamefont {Liu}},\ }\bibfield
  {title} {\bibinfo {title} {Altermagnetism and superconductivity: A short
  historical review},\ }\href {https://arxiv.org/abs/2510.09170} {\  (\bibinfo
  {year} {2026}{\natexlab{a}})},\ \Eprint {https://arxiv.org/abs/2510.09170}
  {arXiv:2510.09170 [cond-mat.supr-con]} \BibitemShut {NoStop}%
\bibitem [{\citenamefont {{Birk Hellenes}}\ \emph {et~al.}(2023)\citenamefont
  {{Birk Hellenes}}, \citenamefont {{Jungwirth}}, \citenamefont
  {{Jaeschke-Ubiergo}}, \citenamefont {{Chakraborty}}, \citenamefont
  {{Sinova}},\ and\ \citenamefont {{{\v{S}}mejkal}}}]{Hellenes2023pwave}%
  \BibitemOpen
  \bibfield  {author} {\bibinfo {author} {\bibfnamefont {A.}~\bibnamefont
  {{Birk Hellenes}}}, \bibinfo {author} {\bibfnamefont {T.}~\bibnamefont
  {{Jungwirth}}}, \bibinfo {author} {\bibfnamefont {R.}~\bibnamefont
  {{Jaeschke-Ubiergo}}}, \bibinfo {author} {\bibfnamefont {A.}~\bibnamefont
  {{Chakraborty}}}, \bibinfo {author} {\bibfnamefont {J.}~\bibnamefont
  {{Sinova}}},\ and\ \bibinfo {author} {\bibfnamefont {L.}~\bibnamefont
  {{{\v{S}}mejkal}}},\ }\bibfield  {title} {\bibinfo {title} {{P-wave
  magnets}},\ }\href {https://doi.org/10.48550/arXiv.2309.01607} {\bibfield
  {journal} {\bibinfo  {journal} {arXiv e-prints}\ ,\ \bibinfo {pages}
  {arXiv:2309.01607}} (\bibinfo {year} {2023})}\BibitemShut {NoStop}%
\bibitem [{\citenamefont {Brekke}\ \emph {et~al.}(2024)\citenamefont {Brekke},
  \citenamefont {Sukhachov}, \citenamefont {Giil}, \citenamefont {Brataas},\
  and\ \citenamefont {Linder}}]{Brekke2024pwave}%
  \BibitemOpen
  \bibfield  {author} {\bibinfo {author} {\bibfnamefont {B.}~\bibnamefont
  {Brekke}}, \bibinfo {author} {\bibfnamefont {P.}~\bibnamefont {Sukhachov}},
  \bibinfo {author} {\bibfnamefont {H.~G.}\ \bibnamefont {Giil}}, \bibinfo
  {author} {\bibfnamefont {A.}~\bibnamefont {Brataas}},\ and\ \bibinfo {author}
  {\bibfnamefont {J.}~\bibnamefont {Linder}},\ }\bibfield  {title} {\bibinfo
  {title} {{Minimal Models and Transport Properties of Unconventional $p$-Wave
  Magnets}},\ }\href {https://doi.org/10.1103/PhysRevLett.133.236703}
  {\bibfield  {journal} {\bibinfo  {journal} {Phys. Rev. Lett.}\ }\textbf
  {\bibinfo {volume} {133}},\ \bibinfo {pages} {236703} (\bibinfo {year}
  {2024})}\BibitemShut {NoStop}%
\bibitem [{\citenamefont {Yu}\ \emph {et~al.}(2025)\citenamefont {Yu},
  \citenamefont {Lyngby}, \citenamefont {Shishidou}, \citenamefont {Roig},
  \citenamefont {Kreisel}, \citenamefont {Weinert}, \citenamefont {Andersen},\
  and\ \citenamefont {Agterberg}}]{Yu2025Odd}%
  \BibitemOpen
  \bibfield  {author} {\bibinfo {author} {\bibfnamefont {Y.}~\bibnamefont
  {Yu}}, \bibinfo {author} {\bibfnamefont {M.~B.}\ \bibnamefont {Lyngby}},
  \bibinfo {author} {\bibfnamefont {T.}~\bibnamefont {Shishidou}}, \bibinfo
  {author} {\bibfnamefont {M.}~\bibnamefont {Roig}}, \bibinfo {author}
  {\bibfnamefont {A.}~\bibnamefont {Kreisel}}, \bibinfo {author} {\bibfnamefont
  {M.}~\bibnamefont {Weinert}}, \bibinfo {author} {\bibfnamefont {B.~M.}\
  \bibnamefont {Andersen}},\ and\ \bibinfo {author} {\bibfnamefont {D.~F.}\
  \bibnamefont {Agterberg}},\ }\bibfield  {title} {\bibinfo {title}
  {{Odd-Parity Magnetism Driven by Antiferromagnetic Exchange}},\ }\href
  {https://doi.org/10.1103/zk69-k6b2} {\bibfield  {journal} {\bibinfo
  {journal} {Phys. Rev. Lett.}\ }\textbf {\bibinfo {volume} {135}},\ \bibinfo
  {pages} {046701} (\bibinfo {year} {2025})}\BibitemShut {NoStop}%
\bibitem [{\citenamefont {Liu}\ \emph {et~al.}(2026{\natexlab{b}})\citenamefont
  {Liu}, \citenamefont {Zhuang}, \citenamefont {Zhu}, \citenamefont {Wu},\ and\
  \citenamefont {Yan}}]{liu2025floquet}%
  \BibitemOpen
  \bibfield  {author} {\bibinfo {author} {\bibfnamefont {D.}~\bibnamefont
  {Liu}}, \bibinfo {author} {\bibfnamefont {Z.-Y.}\ \bibnamefont {Zhuang}},
  \bibinfo {author} {\bibfnamefont {D.}~\bibnamefont {Zhu}}, \bibinfo {author}
  {\bibfnamefont {Z.}~\bibnamefont {Wu}},\ and\ \bibinfo {author}
  {\bibfnamefont {Z.}~\bibnamefont {Yan}},\ }\bibfield  {title} {\bibinfo
  {title} {Light-induced odd-parity altermagnets on dimerized lattices},\
  }\href {https://doi.org/10.1103/wnqs-3djt} {\bibfield  {journal} {\bibinfo
  {journal} {Phys. Rev. B}\ }\textbf {\bibinfo {volume} {113}},\ \bibinfo
  {pages} {L060409} (\bibinfo {year} {2026}{\natexlab{b}})}\BibitemShut
  {NoStop}%
\bibitem [{\citenamefont {Huang}\ \emph {et~al.}(2026)\citenamefont {Huang},
  \citenamefont {Qin}, \citenamefont {Zhan}, \citenamefont {Xu}, \citenamefont
  {Ma},\ and\ \citenamefont {Wang}}]{huang2025oddparityAM}%
  \BibitemOpen
  \bibfield  {author} {\bibinfo {author} {\bibfnamefont {S.}~\bibnamefont
  {Huang}}, \bibinfo {author} {\bibfnamefont {Z.}~\bibnamefont {Qin}}, \bibinfo
  {author} {\bibfnamefont {F.}~\bibnamefont {Zhan}}, \bibinfo {author}
  {\bibfnamefont {D.-H.}\ \bibnamefont {Xu}}, \bibinfo {author} {\bibfnamefont
  {D.-S.}\ \bibnamefont {Ma}},\ and\ \bibinfo {author} {\bibfnamefont
  {R.}~\bibnamefont {Wang}},\ }\bibfield  {title} {\bibinfo {title}
  {Light-induced odd-parity magnetism in conventional antiferromagnetism},\
  }\href {https://doi.org/10.1103/9346-9jpf} {\bibfield  {journal} {\bibinfo
  {journal} {Phys. Rev. Lett.}\ }\textbf {\bibinfo {volume} {136}},\ \bibinfo
  {pages} {126703} (\bibinfo {year} {2026})}\BibitemShut {NoStop}%
\bibitem [{\citenamefont {{Lin}}(2025)}]{lin2025}%
  \BibitemOpen
  \bibfield  {author} {\bibinfo {author} {\bibfnamefont {Y.-P.}\ \bibnamefont
  {{Lin}}},\ }\bibfield  {title} {\bibinfo {title} {{Odd-parity altermagnetism
  through sublattice currents: From Haldane-Hubbard model to general bipartite
  lattices}},\ }\href {https://doi.org/10.48550/arXiv.2503.09602} {\bibfield
  {journal} {\bibinfo  {journal} {arXiv e-prints}\ ,\ \bibinfo {pages}
  {arXiv:2503.09602}} (\bibinfo {year} {2025})}\BibitemShut {NoStop}%
\bibitem [{\citenamefont {{Zeng}}\ \emph {et~al.}(2025)\citenamefont {{Zeng}},
  \citenamefont {{Qin}}, \citenamefont {{Qin}}, \citenamefont {{Feng}},
  \citenamefont {{Xu}},\ and\ \citenamefont {{Wang}}}]{zeng2025}%
  \BibitemOpen
  \bibfield  {author} {\bibinfo {author} {\bibfnamefont {M.}~\bibnamefont
  {{Zeng}}}, \bibinfo {author} {\bibfnamefont {Z.}~\bibnamefont {{Qin}}},
  \bibinfo {author} {\bibfnamefont {L.}~\bibnamefont {{Qin}}}, \bibinfo
  {author} {\bibfnamefont {S.}~\bibnamefont {{Feng}}}, \bibinfo {author}
  {\bibfnamefont {D.-H.}\ \bibnamefont {{Xu}}},\ and\ \bibinfo {author}
  {\bibfnamefont {R.}~\bibnamefont {{Wang}}},\ }\bibfield  {title} {\bibinfo
  {title} {{The electronic and transport properties in the Haldane-Hubbard with
  odd-parity altermagnetism}},\ }\href
  {https://doi.org/10.48550/arXiv.2507.09906} {\bibfield  {journal} {\bibinfo
  {journal} {arXiv e-prints}\ ,\ \bibinfo {pages} {arXiv:2507.09906}} (\bibinfo
  {year} {2025})}\BibitemShut {NoStop}%
\bibitem [{\citenamefont {Zhu}\ \emph {et~al.}(2026)\citenamefont {Zhu},
  \citenamefont {Zhou}, \citenamefont {Wang}, \citenamefont {Wei},\ and\
  \citenamefont {Ruan}}]{zhu2025floquet}%
  \BibitemOpen
  \bibfield  {author} {\bibinfo {author} {\bibfnamefont {T.}~\bibnamefont
  {Zhu}}, \bibinfo {author} {\bibfnamefont {D.}~\bibnamefont {Zhou}}, \bibinfo
  {author} {\bibfnamefont {H.}~\bibnamefont {Wang}}, \bibinfo {author}
  {\bibfnamefont {S.-H.}\ \bibnamefont {Wei}},\ and\ \bibinfo {author}
  {\bibfnamefont {J.}~\bibnamefont {Ruan}},\ }\bibfield  {title} {\bibinfo
  {title} {Floquet odd-parity collinear magnets},\ }\href
  {https://doi.org/10.1103/7ywb-ml2q} {\bibfield  {journal} {\bibinfo
  {journal} {Phys. Rev. Lett.}\ }\textbf {\bibinfo {volume} {136}},\ \bibinfo
  {pages} {126704} (\bibinfo {year} {2026})}\BibitemShut {NoStop}%
\bibitem [{\citenamefont {Li}\ \emph {et~al.}(2026)\citenamefont {Li},
  \citenamefont {Shao},\ and\ \citenamefont {Kovalev}}]{li2025floquet}%
  \BibitemOpen
  \bibfield  {author} {\bibinfo {author} {\bibfnamefont {B.}~\bibnamefont
  {Li}}, \bibinfo {author} {\bibfnamefont {D.-F.}\ \bibnamefont {Shao}},\ and\
  \bibinfo {author} {\bibfnamefont {A.~A.}\ \bibnamefont {Kovalev}},\
  }\bibfield  {title} {\bibinfo {title} {Floquet spin splitting and spin
  generation in antiferromagnets},\ }\href {https://doi.org/10.1103/xzm1-l6yf}
  {\bibfield  {journal} {\bibinfo  {journal} {Phys. Rev. Lett.}\ }\textbf
  {\bibinfo {volume} {136}},\ \bibinfo {pages} {166701} (\bibinfo {year}
  {2026})}\BibitemShut {NoStop}%
\bibitem [{\citenamefont {Zhuang}\ \emph {et~al.}(2025)\citenamefont {Zhuang},
  \citenamefont {Zhu}, \citenamefont {Liu}, \citenamefont {Wu},\ and\
  \citenamefont {Yan}}]{zhuang2025}%
  \BibitemOpen
  \bibfield  {author} {\bibinfo {author} {\bibfnamefont {Z.-Y.}\ \bibnamefont
  {Zhuang}}, \bibinfo {author} {\bibfnamefont {D.}~\bibnamefont {Zhu}},
  \bibinfo {author} {\bibfnamefont {D.}~\bibnamefont {Liu}}, \bibinfo {author}
  {\bibfnamefont {Z.}~\bibnamefont {Wu}},\ and\ \bibinfo {author}
  {\bibfnamefont {Z.}~\bibnamefont {Yan}},\ }\href
  {https://arxiv.org/abs/2508.18361} {\bibinfo {title} {Odd-parity
  altermagnetism originated from orbital orders}} (\bibinfo {year} {2025}),\
  \Eprint {https://arxiv.org/abs/2508.18361} {arXiv:2508.18361
  [cond-mat.mes-hall]} \BibitemShut {NoStop}%
\bibitem [{\citenamefont {Pan}\ \emph {et~al.}(2025)\citenamefont {Pan},
  \citenamefont {Zhou}, \citenamefont {Hu}, \citenamefont {Liu}, \citenamefont
  {Zhou}, \citenamefont {Xiao}, \citenamefont {Yang},\ and\ \citenamefont
  {Sun}}]{Pan2025odd}%
  \BibitemOpen
  \bibfield  {author} {\bibinfo {author} {\bibfnamefont {B.}~\bibnamefont
  {Pan}}, \bibinfo {author} {\bibfnamefont {P.}~\bibnamefont {Zhou}}, \bibinfo
  {author} {\bibfnamefont {Y.}~\bibnamefont {Hu}}, \bibinfo {author}
  {\bibfnamefont {S.}~\bibnamefont {Liu}}, \bibinfo {author} {\bibfnamefont
  {B.}~\bibnamefont {Zhou}}, \bibinfo {author} {\bibfnamefont {H.}~\bibnamefont
  {Xiao}}, \bibinfo {author} {\bibfnamefont {X.}~\bibnamefont {Yang}},\ and\
  \bibinfo {author} {\bibfnamefont {L.}~\bibnamefont {Sun}},\ }\bibfield
  {title} {\bibinfo {title} {Floquet-induced altermagnetic transition in
  $a$-type antiferromagnetic bilayers},\ }\href
  {https://doi.org/10.1103/nn5t-kmln} {\bibfield  {journal} {\bibinfo
  {journal} {Phys. Rev. B}\ }\textbf {\bibinfo {volume} {112}},\ \bibinfo
  {pages} {224430} (\bibinfo {year} {2025})}\BibitemShut {NoStop}%
\bibitem [{\citenamefont {Liu}\ \emph {et~al.}(2022)\citenamefont {Liu},
  \citenamefont {Li}, \citenamefont {Han}, \citenamefont {Wan},\ and\
  \citenamefont {Liu}}]{Liu2022AM}%
  \BibitemOpen
  \bibfield  {author} {\bibinfo {author} {\bibfnamefont {P.}~\bibnamefont
  {Liu}}, \bibinfo {author} {\bibfnamefont {J.}~\bibnamefont {Li}}, \bibinfo
  {author} {\bibfnamefont {J.}~\bibnamefont {Han}}, \bibinfo {author}
  {\bibfnamefont {X.}~\bibnamefont {Wan}},\ and\ \bibinfo {author}
  {\bibfnamefont {Q.}~\bibnamefont {Liu}},\ }\bibfield  {title} {\bibinfo
  {title} {{Spin-Group Symmetry in Magnetic Materials with Negligible
  Spin-Orbit Coupling}},\ }\href {https://doi.org/10.1103/PhysRevX.12.021016}
  {\bibfield  {journal} {\bibinfo  {journal} {Phys. Rev. X}\ }\textbf {\bibinfo
  {volume} {12}},\ \bibinfo {pages} {021016} (\bibinfo {year}
  {2022})}\BibitemShut {NoStop}%
\bibitem [{\citenamefont {Chen}\ \emph {et~al.}(2024)\citenamefont {Chen},
  \citenamefont {Ren}, \citenamefont {Zhu}, \citenamefont {Yu}, \citenamefont
  {Zhang}, \citenamefont {Liu}, \citenamefont {Li}, \citenamefont {Liu},
  \citenamefont {Li},\ and\ \citenamefont {Liu}}]{Liu2024AMPRX}%
  \BibitemOpen
  \bibfield  {author} {\bibinfo {author} {\bibfnamefont {X.}~\bibnamefont
  {Chen}}, \bibinfo {author} {\bibfnamefont {J.}~\bibnamefont {Ren}}, \bibinfo
  {author} {\bibfnamefont {Y.}~\bibnamefont {Zhu}}, \bibinfo {author}
  {\bibfnamefont {Y.}~\bibnamefont {Yu}}, \bibinfo {author} {\bibfnamefont
  {A.}~\bibnamefont {Zhang}}, \bibinfo {author} {\bibfnamefont
  {P.}~\bibnamefont {Liu}}, \bibinfo {author} {\bibfnamefont {J.}~\bibnamefont
  {Li}}, \bibinfo {author} {\bibfnamefont {Y.}~\bibnamefont {Liu}}, \bibinfo
  {author} {\bibfnamefont {C.}~\bibnamefont {Li}},\ and\ \bibinfo {author}
  {\bibfnamefont {Q.}~\bibnamefont {Liu}},\ }\bibfield  {title} {\bibinfo
  {title} {{Enumeration and Representation Theory of Spin Space Groups}},\
  }\href {https://doi.org/10.1103/PhysRevX.14.031038} {\bibfield  {journal}
  {\bibinfo  {journal} {Phys. Rev. X}\ }\textbf {\bibinfo {volume} {14}},\
  \bibinfo {pages} {031038} (\bibinfo {year} {2024})}\BibitemShut {NoStop}%
\bibitem [{\citenamefont {Jiang}\ \emph {et~al.}(2024)\citenamefont {Jiang},
  \citenamefont {Song}, \citenamefont {Zhu}, \citenamefont {Fang},
  \citenamefont {Weng}, \citenamefont {Liu}, \citenamefont {Yang},\ and\
  \citenamefont {Fang}}]{Jiang2024SSG}%
  \BibitemOpen
  \bibfield  {author} {\bibinfo {author} {\bibfnamefont {Y.}~\bibnamefont
  {Jiang}}, \bibinfo {author} {\bibfnamefont {Z.}~\bibnamefont {Song}},
  \bibinfo {author} {\bibfnamefont {T.}~\bibnamefont {Zhu}}, \bibinfo {author}
  {\bibfnamefont {Z.}~\bibnamefont {Fang}}, \bibinfo {author} {\bibfnamefont
  {H.}~\bibnamefont {Weng}}, \bibinfo {author} {\bibfnamefont {Z.-X.}\
  \bibnamefont {Liu}}, \bibinfo {author} {\bibfnamefont {J.}~\bibnamefont
  {Yang}},\ and\ \bibinfo {author} {\bibfnamefont {C.}~\bibnamefont {Fang}},\
  }\bibfield  {title} {\bibinfo {title} {Enumeration of spin-space groups:
  Toward a complete description of symmetries of magnetic orders},\ }\href
  {https://doi.org/10.1103/PhysRevX.14.031039} {\bibfield  {journal} {\bibinfo
  {journal} {Phys. Rev. X}\ }\textbf {\bibinfo {volume} {14}},\ \bibinfo
  {pages} {031039} (\bibinfo {year} {2024})}\BibitemShut {NoStop}%
\bibitem [{\citenamefont {Xiao}\ \emph {et~al.}(2024)\citenamefont {Xiao},
  \citenamefont {Zhao}, \citenamefont {Li}, \citenamefont {Shindou},\ and\
  \citenamefont {Song}}]{Xiao2024SSG}%
  \BibitemOpen
  \bibfield  {author} {\bibinfo {author} {\bibfnamefont {Z.}~\bibnamefont
  {Xiao}}, \bibinfo {author} {\bibfnamefont {J.}~\bibnamefont {Zhao}}, \bibinfo
  {author} {\bibfnamefont {Y.}~\bibnamefont {Li}}, \bibinfo {author}
  {\bibfnamefont {R.}~\bibnamefont {Shindou}},\ and\ \bibinfo {author}
  {\bibfnamefont {Z.-D.}\ \bibnamefont {Song}},\ }\bibfield  {title} {\bibinfo
  {title} {{Spin Space Groups: Full Classification and Applications}},\ }\href
  {https://doi.org/10.1103/PhysRevX.14.031037} {\bibfield  {journal} {\bibinfo
  {journal} {Phys. Rev. X}\ }\textbf {\bibinfo {volume} {14}},\ \bibinfo
  {pages} {031037} (\bibinfo {year} {2024})}\BibitemShut {NoStop}%
\bibitem [{\citenamefont {Liu}\ \emph {et~al.}(2026{\natexlab{c}})\citenamefont
  {Liu}, \citenamefont {Chen}, \citenamefont {Yu}, \citenamefont {Etxebarria},
  \citenamefont {Perez-Mato},\ and\ \citenamefont {Liu}}]{liu2026OSSG}%
  \BibitemOpen
  \bibfield  {author} {\bibinfo {author} {\bibfnamefont {Y.}~\bibnamefont
  {Liu}}, \bibinfo {author} {\bibfnamefont {X.}~\bibnamefont {Chen}}, \bibinfo
  {author} {\bibfnamefont {Y.}~\bibnamefont {Yu}}, \bibinfo {author}
  {\bibfnamefont {J.}~\bibnamefont {Etxebarria}}, \bibinfo {author}
  {\bibfnamefont {J.~M.}\ \bibnamefont {Perez-Mato}},\ and\ \bibinfo {author}
  {\bibfnamefont {Q.}~\bibnamefont {Liu}},\ }\bibfield  {title} {\bibinfo
  {title} {Symmetry classification of magnetic orders using oriented spin space
  groups},\ }\href {https://doi.org/10.1038/s41586-026-10401-1} {\bibfield
  {journal} {\bibinfo  {journal} {Nature}\ }\textbf {\bibinfo {volume} {652}},\
  \bibinfo {pages} {869} (\bibinfo {year} {2026}{\natexlab{c}})}\BibitemShut
  {NoStop}%
\bibitem [{\citenamefont {Zaanen}\ and\ \citenamefont
  {Gunnarsson}(1989)}]{Zaanen1989stripe}%
  \BibitemOpen
  \bibfield  {author} {\bibinfo {author} {\bibfnamefont {J.}~\bibnamefont
  {Zaanen}}\ and\ \bibinfo {author} {\bibfnamefont {O.}~\bibnamefont
  {Gunnarsson}},\ }\bibfield  {title} {\bibinfo {title} {Charged magnetic
  domain lines and the magnetism of high-${T}_{c}$ oxides},\ }\href
  {https://doi.org/10.1103/PhysRevB.40.7391} {\bibfield  {journal} {\bibinfo
  {journal} {Phys. Rev. B}\ }\textbf {\bibinfo {volume} {40}},\ \bibinfo
  {pages} {7391(R)} (\bibinfo {year} {1989})}\BibitemShut {NoStop}%
\bibitem [{\citenamefont {Machida}(1989)}]{Machida1989}%
  \BibitemOpen
  \bibfield  {author} {\bibinfo {author} {\bibfnamefont {K.}~\bibnamefont
  {Machida}},\ }\bibfield  {title} {\bibinfo {title} {Magnetism in la2cuo4
  based compounds},\ }\href
  {https://doi.org/https://doi.org/10.1016/0921-4534(89)90316-X} {\bibfield
  {journal} {\bibinfo  {journal} {Physica C: Superconductivity}\ }\textbf
  {\bibinfo {volume} {158}},\ \bibinfo {pages} {192} (\bibinfo {year}
  {1989})}\BibitemShut {NoStop}%
\bibitem [{\citenamefont {Kato}\ \emph {et~al.}(1990)\citenamefont {Kato},
  \citenamefont {Machida}, \citenamefont {Nakanishi},\ and\ \citenamefont
  {Fujita}}]{Kato1990}%
  \BibitemOpen
  \bibfield  {author} {\bibinfo {author} {\bibfnamefont {M.}~\bibnamefont
  {Kato}}, \bibinfo {author} {\bibfnamefont {K.}~\bibnamefont {Machida}},
  \bibinfo {author} {\bibfnamefont {H.}~\bibnamefont {Nakanishi}},\ and\
  \bibinfo {author} {\bibfnamefont {M.}~\bibnamefont {Fujita}},\ }\bibfield
  {title} {\bibinfo {title} {Soliton lattice modulation of incommensurate spin
  density wave in two dimensional hubbard model -a mean field study-},\ }\href
  {https://doi.org/10.1143/JPSJ.59.1047} {\bibfield  {journal} {\bibinfo
  {journal} {Journal of the Physical Society of Japan}\ }\textbf {\bibinfo
  {volume} {59}},\ \bibinfo {pages} {1047} (\bibinfo {year}
  {1990})}\BibitemShut {NoStop}%
\bibitem [{\citenamefont {H{\"u}cker}(2012)}]{hucker2012stripe}%
  \BibitemOpen
  \bibfield  {author} {\bibinfo {author} {\bibfnamefont {M.}~\bibnamefont
  {H{\"u}cker}},\ }\bibfield  {title} {\bibinfo {title} {Structural aspects of
  materials with static stripe order},\ }\href
  {https://doi.org/10.1016/j.physc.2012.04.035} {\bibfield  {journal} {\bibinfo
   {journal} {Physica C: Superconductivity}\ }\textbf {\bibinfo {volume}
  {481}},\ \bibinfo {pages} {3} (\bibinfo {year} {2012})}\BibitemShut {NoStop}%
\bibitem [{\citenamefont {Tranquada}\ \emph {et~al.}(1995)\citenamefont
  {Tranquada}, \citenamefont {Sternlieb}, \citenamefont {Axe}, \citenamefont
  {Nakamura},\ and\ \citenamefont {Uchida}}]{tranquada1995stripe}%
  \BibitemOpen
  \bibfield  {author} {\bibinfo {author} {\bibfnamefont {J.}~\bibnamefont
  {Tranquada}}, \bibinfo {author} {\bibfnamefont {B.}~\bibnamefont
  {Sternlieb}}, \bibinfo {author} {\bibfnamefont {J.}~\bibnamefont {Axe}},
  \bibinfo {author} {\bibfnamefont {Y.}~\bibnamefont {Nakamura}},\ and\
  \bibinfo {author} {\bibfnamefont {S.-i.}\ \bibnamefont {Uchida}},\ }\bibfield
   {title} {\bibinfo {title} {Evidence for stripe correlations of spins and
  holes in copper oxide superconductors},\ }\href
  {https://doi.org/doi.org/10.1038/375561a0} {\bibfield  {journal} {\bibinfo
  {journal} {nature}\ }\textbf {\bibinfo {volume} {375}},\ \bibinfo {pages}
  {561} (\bibinfo {year} {1995})}\BibitemShut {NoStop}%
\bibitem [{\citenamefont {Kivelson}\ \emph {et~al.}(2003)\citenamefont
  {Kivelson}, \citenamefont {Bindloss}, \citenamefont {Fradkin}, \citenamefont
  {Oganesyan}, \citenamefont {Tranquada}, \citenamefont {Kapitulnik},\ and\
  \citenamefont {Howald}}]{Kivelson2003stripe}%
  \BibitemOpen
  \bibfield  {author} {\bibinfo {author} {\bibfnamefont {S.~A.}\ \bibnamefont
  {Kivelson}}, \bibinfo {author} {\bibfnamefont {I.~P.}\ \bibnamefont
  {Bindloss}}, \bibinfo {author} {\bibfnamefont {E.}~\bibnamefont {Fradkin}},
  \bibinfo {author} {\bibfnamefont {V.}~\bibnamefont {Oganesyan}}, \bibinfo
  {author} {\bibfnamefont {J.~M.}\ \bibnamefont {Tranquada}}, \bibinfo {author}
  {\bibfnamefont {A.}~\bibnamefont {Kapitulnik}},\ and\ \bibinfo {author}
  {\bibfnamefont {C.}~\bibnamefont {Howald}},\ }\bibfield  {title} {\bibinfo
  {title} {How to detect fluctuating stripes in the high-temperature
  superconductors},\ }\href {https://doi.org/10.1103/RevModPhys.75.1201}
  {\bibfield  {journal} {\bibinfo  {journal} {Rev. Mod. Phys.}\ }\textbf
  {\bibinfo {volume} {75}},\ \bibinfo {pages} {1201} (\bibinfo {year}
  {2003})}\BibitemShut {NoStop}%
\bibitem [{\citenamefont {Missiaen}\ \emph {et~al.}(2025)\citenamefont
  {Missiaen}, \citenamefont {Mayaffre}, \citenamefont {Kr\"amer}, \citenamefont
  {Zhao}, \citenamefont {Zhou}, \citenamefont {Wu}, \citenamefont {Chen},
  \citenamefont {Pyon}, \citenamefont {Takayama}, \citenamefont {Takagi},
  \citenamefont {LeBoeuf},\ and\ \citenamefont {Julien}}]{Missiaen2025stripe}%
  \BibitemOpen
  \bibfield  {author} {\bibinfo {author} {\bibfnamefont {A.}~\bibnamefont
  {Missiaen}}, \bibinfo {author} {\bibfnamefont {H.}~\bibnamefont {Mayaffre}},
  \bibinfo {author} {\bibfnamefont {S.}~\bibnamefont {Kr\"amer}}, \bibinfo
  {author} {\bibfnamefont {D.}~\bibnamefont {Zhao}}, \bibinfo {author}
  {\bibfnamefont {Y.}~\bibnamefont {Zhou}}, \bibinfo {author} {\bibfnamefont
  {T.}~\bibnamefont {Wu}}, \bibinfo {author} {\bibfnamefont {X.}~\bibnamefont
  {Chen}}, \bibinfo {author} {\bibfnamefont {S.}~\bibnamefont {Pyon}}, \bibinfo
  {author} {\bibfnamefont {T.}~\bibnamefont {Takayama}}, \bibinfo {author}
  {\bibfnamefont {H.}~\bibnamefont {Takagi}}, \bibinfo {author} {\bibfnamefont
  {D.}~\bibnamefont {LeBoeuf}},\ and\ \bibinfo {author} {\bibfnamefont {M.-H.}\
  \bibnamefont {Julien}},\ }\bibfield  {title} {\bibinfo {title} {Spin-stripe
  order tied to the pseudogap phase in
  ${\mathrm{la}}_{1.8\ensuremath{-}x}{\mathrm{eu}}_{0.2}{\mathrm{sr}}_{x}{\mathrm{cuo}}_{4}$},\
  }\href {https://doi.org/10.1103/PhysRevX.15.021010} {\bibfield  {journal}
  {\bibinfo  {journal} {Phys. Rev. X}\ }\textbf {\bibinfo {volume} {15}},\
  \bibinfo {pages} {021010} (\bibinfo {year} {2025})}\BibitemShut {NoStop}%
\bibitem [{\citenamefont {Fang}\ \emph {et~al.}(2008)\citenamefont {Fang},
  \citenamefont {Yao}, \citenamefont {Tsai}, \citenamefont {Hu},\ and\
  \citenamefont {Kivelson}}]{Chen2008nematic}%
  \BibitemOpen
  \bibfield  {author} {\bibinfo {author} {\bibfnamefont {C.}~\bibnamefont
  {Fang}}, \bibinfo {author} {\bibfnamefont {H.}~\bibnamefont {Yao}}, \bibinfo
  {author} {\bibfnamefont {W.-F.}\ \bibnamefont {Tsai}}, \bibinfo {author}
  {\bibfnamefont {J.}~\bibnamefont {Hu}},\ and\ \bibinfo {author}
  {\bibfnamefont {S.~A.}\ \bibnamefont {Kivelson}},\ }\bibfield  {title}
  {\bibinfo {title} {Theory of electron nematic order in lafeaso},\ }\href
  {https://doi.org/10.1103/PhysRevB.77.224509} {\bibfield  {journal} {\bibinfo
  {journal} {Phys. Rev. B}\ }\textbf {\bibinfo {volume} {77}},\ \bibinfo
  {pages} {224509} (\bibinfo {year} {2008})}\BibitemShut {NoStop}%
\bibitem [{\citenamefont {Xu}\ \emph {et~al.}(2008)\citenamefont {Xu},
  \citenamefont {M\"uller},\ and\ \citenamefont {Sachdev}}]{Xu2008stripe}%
  \BibitemOpen
  \bibfield  {author} {\bibinfo {author} {\bibfnamefont {C.}~\bibnamefont
  {Xu}}, \bibinfo {author} {\bibfnamefont {M.}~\bibnamefont {M\"uller}},\ and\
  \bibinfo {author} {\bibfnamefont {S.}~\bibnamefont {Sachdev}},\ }\bibfield
  {title} {\bibinfo {title} {Ising and spin orders in the iron-based
  superconductors},\ }\href {https://doi.org/10.1103/PhysRevB.78.020501}
  {\bibfield  {journal} {\bibinfo  {journal} {Phys. Rev. B}\ }\textbf {\bibinfo
  {volume} {78}},\ \bibinfo {pages} {020501(R)} (\bibinfo {year}
  {2008})}\BibitemShut {NoStop}%
\bibitem [{\citenamefont {Fernandes}\ \emph {et~al.}(2012)\citenamefont
  {Fernandes}, \citenamefont {Chubukov}, \citenamefont {Knolle}, \citenamefont
  {Eremin},\ and\ \citenamefont {Schmalian}}]{Fernandes2012nematic}%
  \BibitemOpen
  \bibfield  {author} {\bibinfo {author} {\bibfnamefont {R.~M.}\ \bibnamefont
  {Fernandes}}, \bibinfo {author} {\bibfnamefont {A.~V.}\ \bibnamefont
  {Chubukov}}, \bibinfo {author} {\bibfnamefont {J.}~\bibnamefont {Knolle}},
  \bibinfo {author} {\bibfnamefont {I.}~\bibnamefont {Eremin}},\ and\ \bibinfo
  {author} {\bibfnamefont {J.}~\bibnamefont {Schmalian}},\ }\bibfield  {title}
  {\bibinfo {title} {Preemptive nematic order, pseudogap, and orbital order in
  the iron pnictides},\ }\href {https://doi.org/10.1103/PhysRevB.85.024534}
  {\bibfield  {journal} {\bibinfo  {journal} {Phys. Rev. B}\ }\textbf {\bibinfo
  {volume} {85}},\ \bibinfo {pages} {024534} (\bibinfo {year}
  {2012})}\BibitemShut {NoStop}%
\bibitem [{\citenamefont {Achkar}\ \emph {et~al.}(2016)\citenamefont {Achkar},
  \citenamefont {Zwiebler}, \citenamefont {McMahon}, \citenamefont {He},
  \citenamefont {Sutarto}, \citenamefont {Djianto}, \citenamefont {Hao},
  \citenamefont {Gingras}, \citenamefont {Hücker}, \citenamefont {Gu},
  \citenamefont {Revcolevschi}, \citenamefont {Zhang}, \citenamefont {Kim},
  \citenamefont {Geck},\ and\ \citenamefont {Hawthorn}}]{Achkar2016stripe}%
  \BibitemOpen
  \bibfield  {author} {\bibinfo {author} {\bibfnamefont {A.~J.}\ \bibnamefont
  {Achkar}}, \bibinfo {author} {\bibfnamefont {M.}~\bibnamefont {Zwiebler}},
  \bibinfo {author} {\bibfnamefont {C.}~\bibnamefont {McMahon}}, \bibinfo
  {author} {\bibfnamefont {F.}~\bibnamefont {He}}, \bibinfo {author}
  {\bibfnamefont {R.}~\bibnamefont {Sutarto}}, \bibinfo {author} {\bibfnamefont
  {I.}~\bibnamefont {Djianto}}, \bibinfo {author} {\bibfnamefont
  {Z.}~\bibnamefont {Hao}}, \bibinfo {author} {\bibfnamefont {M.~J.~P.}\
  \bibnamefont {Gingras}}, \bibinfo {author} {\bibfnamefont {M.}~\bibnamefont
  {Hücker}}, \bibinfo {author} {\bibfnamefont {G.~D.}\ \bibnamefont {Gu}},
  \bibinfo {author} {\bibfnamefont {A.}~\bibnamefont {Revcolevschi}}, \bibinfo
  {author} {\bibfnamefont {H.}~\bibnamefont {Zhang}}, \bibinfo {author}
  {\bibfnamefont {Y.-J.}\ \bibnamefont {Kim}}, \bibinfo {author} {\bibfnamefont
  {J.}~\bibnamefont {Geck}},\ and\ \bibinfo {author} {\bibfnamefont {D.~G.}\
  \bibnamefont {Hawthorn}},\ }\bibfield  {title} {\bibinfo {title} {Nematicity
  in stripe-ordered cuprates probed via resonant x-ray scattering},\ }\href
  {https://doi.org/10.1126/science.aad1824} {\bibfield  {journal} {\bibinfo
  {journal} {Science}\ }\textbf {\bibinfo {volume} {351}},\ \bibinfo {pages}
  {576} (\bibinfo {year} {2016})}\BibitemShut {NoStop}%
\bibitem [{\citenamefont {Leeb}\ \emph {et~al.}(2024)\citenamefont {Leeb},
  \citenamefont {Mook}, \citenamefont {\ifmmode~\check{S}\else
  \v{S}\fi{}mejkal},\ and\ \citenamefont {Knolle}}]{Leeb2024AM}%
  \BibitemOpen
  \bibfield  {author} {\bibinfo {author} {\bibfnamefont {V.}~\bibnamefont
  {Leeb}}, \bibinfo {author} {\bibfnamefont {A.}~\bibnamefont {Mook}}, \bibinfo
  {author} {\bibfnamefont {L.}~\bibnamefont {\ifmmode~\check{S}\else
  \v{S}\fi{}mejkal}},\ and\ \bibinfo {author} {\bibfnamefont {J.}~\bibnamefont
  {Knolle}},\ }\bibfield  {title} {\bibinfo {title} {{Spontaneous Formation of
  Altermagnetism from Orbital Ordering}},\ }\href
  {https://doi.org/10.1103/PhysRevLett.132.236701} {\bibfield  {journal}
  {\bibinfo  {journal} {Phys. Rev. Lett.}\ }\textbf {\bibinfo {volume} {132}},\
  \bibinfo {pages} {236701} (\bibinfo {year} {2024})}\BibitemShut {NoStop}%
\bibitem [{\citenamefont {Vila}\ \emph {et~al.}(2025)\citenamefont {Vila},
  \citenamefont {Sunko},\ and\ \citenamefont {Moore}}]{Vila2025orbitalorder}%
  \BibitemOpen
  \bibfield  {author} {\bibinfo {author} {\bibfnamefont {M.}~\bibnamefont
  {Vila}}, \bibinfo {author} {\bibfnamefont {V.}~\bibnamefont {Sunko}},\ and\
  \bibinfo {author} {\bibfnamefont {J.~E.}\ \bibnamefont {Moore}},\ }\bibfield
  {title} {\bibinfo {title} {Orbital-spin locking and its optical signatures in
  altermagnets},\ }\href {https://doi.org/10.1103/bzzy-ngcs} {\bibfield
  {journal} {\bibinfo  {journal} {Phys. Rev. B}\ }\textbf {\bibinfo {volume}
  {112}},\ \bibinfo {pages} {L020401} (\bibinfo {year} {2025})}\BibitemShut
  {NoStop}%
\bibitem [{\citenamefont {Meier}\ \emph {et~al.}(2026)\citenamefont {Meier},
  \citenamefont {Carta}, \citenamefont {Ederer},\ and\ \citenamefont
  {Cano}}]{Meier2026}%
  \BibitemOpen
  \bibfield  {author} {\bibinfo {author} {\bibfnamefont {Q.~N.}\ \bibnamefont
  {Meier}}, \bibinfo {author} {\bibfnamefont {A.}~\bibnamefont {Carta}},
  \bibinfo {author} {\bibfnamefont {C.}~\bibnamefont {Ederer}},\ and\ \bibinfo
  {author} {\bibfnamefont {A.}~\bibnamefont {Cano}},\ }\bibfield  {title}
  {\bibinfo {title} {Net and compensated altermagnetism from staggered orbital
  order: Layer-dependent spin splitting in
  ${\mathrm{sr}}_{n+1}{\mathrm{cr}}_{n}{\mathrm{o}}_{3n+1}$},\ }\href
  {https://doi.org/10.1103/mm8t-82q4} {\bibfield  {journal} {\bibinfo
  {journal} {Phys. Rev. Lett.}\ }\textbf {\bibinfo {volume} {136}},\ \bibinfo
  {pages} {116705} (\bibinfo {year} {2026})}\BibitemShut {NoStop}%
\bibitem [{\citenamefont {Camerano}\ \emph {et~al.}(2025)\citenamefont
  {Camerano}, \citenamefont {Fumega}, \citenamefont {Lado}, \citenamefont
  {Stroppa},\ and\ \citenamefont {Profeta}}]{Camerano2025}%
  \BibitemOpen
  \bibfield  {author} {\bibinfo {author} {\bibfnamefont {L.}~\bibnamefont
  {Camerano}}, \bibinfo {author} {\bibfnamefont {A.~O.}\ \bibnamefont
  {Fumega}}, \bibinfo {author} {\bibfnamefont {J.~L.}\ \bibnamefont {Lado}},
  \bibinfo {author} {\bibfnamefont {A.}~\bibnamefont {Stroppa}},\ and\ \bibinfo
  {author} {\bibfnamefont {G.}~\bibnamefont {Profeta}},\ }\bibfield  {title}
  {\bibinfo {title} {Multiferroic nematic d-wave altermagnetism driven by
  orbital-order on the honeycomb lattice}\ }\href
  {https://doi.org/10.1038/s41699-025-00599-5} {10.1038/s41699-025-00599-5}
  (\bibinfo {year} {2025})\BibitemShut {NoStop}%
\bibitem [{\citenamefont {Matsuda}\ \emph {et~al.}(2025)\citenamefont
  {Matsuda}, \citenamefont {Watanabe},\ and\ \citenamefont
  {Arita}}]{Matsuda2025}%
  \BibitemOpen
  \bibfield  {author} {\bibinfo {author} {\bibfnamefont {J.}~\bibnamefont
  {Matsuda}}, \bibinfo {author} {\bibfnamefont {H.}~\bibnamefont {Watanabe}},\
  and\ \bibinfo {author} {\bibfnamefont {R.}~\bibnamefont {Arita}},\ }\bibfield
   {title} {\bibinfo {title} {Multiferroic collinear antiferromagnets with
  hidden altermagnetic spin splitting},\ }\href
  {https://doi.org/10.1103/vgcs-bn8g} {\bibfield  {journal} {\bibinfo
  {journal} {Phys. Rev. Lett.}\ }\textbf {\bibinfo {volume} {134}},\ \bibinfo
  {pages} {226703} (\bibinfo {year} {2025})}\BibitemShut {NoStop}%
\bibitem [{sup()}]{supplemental}%
  \BibitemOpen
  \href@noop {} {}\bibinfo {howpublished} {See the Supplemental Material for
  the stripe anti-altermagnets, spontaneous stripe-order altermagnetism, and
  spin-current response.}\BibitemShut {Stop}%
\bibitem [{\citenamefont {Xiao}\ \emph {et~al.}(2010)\citenamefont {Xiao},
  \citenamefont {Chang},\ and\ \citenamefont {Niu}}]{Xiao2010review}%
  \BibitemOpen
  \bibfield  {author} {\bibinfo {author} {\bibfnamefont {D.}~\bibnamefont
  {Xiao}}, \bibinfo {author} {\bibfnamefont {M.-C.}\ \bibnamefont {Chang}},\
  and\ \bibinfo {author} {\bibfnamefont {Q.}~\bibnamefont {Niu}},\ }\bibfield
  {title} {\bibinfo {title} {Berry phase effects on electronic properties},\
  }\href {https://doi.org/10.1103/RevModPhys.82.1959} {\bibfield  {journal}
  {\bibinfo  {journal} {Rev. Mod. Phys.}\ }\textbf {\bibinfo {volume} {82}},\
  \bibinfo {pages} {1959} (\bibinfo {year} {2010})}\BibitemShut {NoStop}%
\bibitem [{\citenamefont {\ifmmode~\check{Z}\else \v{Z}\fi{}elezn\'y}\ \emph
  {et~al.}(2017)\citenamefont {\ifmmode~\check{Z}\else \v{Z}\fi{}elezn\'y},
  \citenamefont {Zhang}, \citenamefont {Felser},\ and\ \citenamefont
  {Yan}}]{Yan2017PRLMn3Ir}%
  \BibitemOpen
  \bibfield  {author} {\bibinfo {author} {\bibfnamefont {J.}~\bibnamefont
  {\ifmmode~\check{Z}\else \v{Z}\fi{}elezn\'y}}, \bibinfo {author}
  {\bibfnamefont {Y.}~\bibnamefont {Zhang}}, \bibinfo {author} {\bibfnamefont
  {C.}~\bibnamefont {Felser}},\ and\ \bibinfo {author} {\bibfnamefont
  {B.}~\bibnamefont {Yan}},\ }\bibfield  {title} {\bibinfo {title}
  {Spin-polarized current in noncollinear antiferromagnets},\ }\href
  {https://doi.org/10.1103/PhysRevLett.119.187204} {\bibfield  {journal}
  {\bibinfo  {journal} {Phys. Rev. Lett.}\ }\textbf {\bibinfo {volume} {119}},\
  \bibinfo {pages} {187204} (\bibinfo {year} {2017})}\BibitemShut {NoStop}%
\bibitem [{\citenamefont {Lenz}\ \emph {et~al.}(2024)\citenamefont {Lenz},
  \citenamefont {Fabrizio},\ and\ \citenamefont {Casula}}]{lenz2024}%
  \BibitemOpen
  \bibfield  {author} {\bibinfo {author} {\bibfnamefont {B.}~\bibnamefont
  {Lenz}}, \bibinfo {author} {\bibfnamefont {M.}~\bibnamefont {Fabrizio}},\
  and\ \bibinfo {author} {\bibfnamefont {M.}~\bibnamefont {Casula}},\
  }\bibfield  {title} {\bibinfo {title} {Order from disorder phenomena in
  bacos2},\ }\href {https://doi.org/10.1038/s42005-023-01514-4} {\bibfield
  {journal} {\bibinfo  {journal} {Communications Physics}\ }\textbf {\bibinfo
  {volume} {7}},\ \bibinfo {pages} {35} (\bibinfo {year} {2024})}\BibitemShut
  {NoStop}%
\bibitem [{\citenamefont {Yamagata}\ \emph {et~al.}(2017)\citenamefont
  {Yamagata}, \citenamefont {Endo}, \citenamefont {Inoue},\ and\ \citenamefont
  {Koyama}}]{yamagata2017}%
  \BibitemOpen
  \bibfield  {author} {\bibinfo {author} {\bibfnamefont {M.}~\bibnamefont
  {Yamagata}}, \bibinfo {author} {\bibfnamefont {T.}~\bibnamefont {Endo}},
  \bibinfo {author} {\bibfnamefont {Y.}~\bibnamefont {Inoue}},\ and\ \bibinfo
  {author} {\bibfnamefont {Y.}~\bibnamefont {Koyama}},\ }\bibfield  {title}
  {\bibinfo {title} {Features of highly correlated electronic states in the
  simple perovskite manganite sr$_{1- x}$ sm$_x$ mno$_3$ with 0.15$\leq$
  x$\leq$ 0.50},\ }\href {https://doi.org/10.7566/JPSJ.86.054705} {\bibfield
  {journal} {\bibinfo  {journal} {journal of the physical society of japan}\
  }\textbf {\bibinfo {volume} {86}},\ \bibinfo {pages} {054705} (\bibinfo
  {year} {2017})}\BibitemShut {NoStop}%
\bibitem [{\citenamefont {Zhuang}\ \emph {et~al.}(2026)\citenamefont {Zhuang},
  \citenamefont {Hu}, \citenamefont {Zhang}, \citenamefont {Hu},\ and\
  \citenamefont {Yan}}]{zhuang2026mixed}%
  \BibitemOpen
  \bibfield  {author} {\bibinfo {author} {\bibfnamefont {Z.-Y.}\ \bibnamefont
  {Zhuang}}, \bibinfo {author} {\bibfnamefont {J.-X.}\ \bibnamefont {Hu}},
  \bibinfo {author} {\bibfnamefont {S.-B.}\ \bibnamefont {Zhang}}, \bibinfo
  {author} {\bibfnamefont {L.-H.}\ \bibnamefont {Hu}},\ and\ \bibinfo {author}
  {\bibfnamefont {Z.}~\bibnamefont {Yan}},\ }\href
  {https://arxiv.org/abs/2605.05205} {\bibinfo {title} {Mixed-parity
  altermagnetism in collinear spin-orbital magnets}} (\bibinfo {year} {2026}),\
  \Eprint {https://arxiv.org/abs/2605.05205} {arXiv:2605.05205
  [cond-mat.mes-hall]} \BibitemShut {NoStop}%
\bibitem [{\citenamefont {Roig}\ \emph {et~al.}(2024)\citenamefont {Roig},
  \citenamefont {Kreisel}, \citenamefont {Yu}, \citenamefont {Andersen},\ and\
  \citenamefont {Agterberg}}]{Roig2024AM}%
  \BibitemOpen
  \bibfield  {author} {\bibinfo {author} {\bibfnamefont {M.}~\bibnamefont
  {Roig}}, \bibinfo {author} {\bibfnamefont {A.}~\bibnamefont {Kreisel}},
  \bibinfo {author} {\bibfnamefont {Y.}~\bibnamefont {Yu}}, \bibinfo {author}
  {\bibfnamefont {B.~M.}\ \bibnamefont {Andersen}},\ and\ \bibinfo {author}
  {\bibfnamefont {D.~F.}\ \bibnamefont {Agterberg}},\ }\bibfield  {title}
  {\bibinfo {title} {Minimal models for altermagnetism},\ }\href
  {https://doi.org/10.1103/PhysRevB.110.144412} {\bibfield  {journal} {\bibinfo
   {journal} {Phys. Rev. B}\ }\textbf {\bibinfo {volume} {110}},\ \bibinfo
  {pages} {144412} (\bibinfo {year} {2024})}\BibitemShut {NoStop}%
\end{thebibliography}%

\appendix 
\clearpage
\begin{center}
	\large \textbf{End Matter} 
\end{center}
\renewcommand{\thefigure}{E\arabic{figure}}
\renewcommand{\theequation}{E\arabic{equation}}
\renewcommand{\thetable}{E\arabic{table}}
\setcounter{equation}{0}
\setcounter{table}{0}
\setcounter{figure}{0}

{\it \color{blue}Spin degeneracy from $[C_{2}\Vert C_{2z}]$ and spinless time reversal.--} In the two-dimensional stripe geometry, among the crystallographic operations $[C_2\Vert C_n]$ with $n=1,2,3,4,6$, the cases $n=1,3$ are incompatible with a finite collinear moment because an odd power generates pure spin reversal $[C_2\Vert E]$, where $E$ is the real-space identity. The $n=4,6$ operations rotate the ferromagnetic chains away from the fixed stripe direction and therefore do not preserve a single ordering vector $\bq$ for stripe order. The only crystallographic spin-reversing rotation that can remain is thus $[C_{2}\Vert C_{2z}]$.

In the nonrelativistic collinear limit considered here, and in the absence of orbital time-reversal breaking, the Hamiltonian also preserves the spinless time-reversal symmetry $[\bar{C}_{2}\Vert\mathcal{T}]$. This operation combines a $180^{\circ}$ spin rotation with time reversal, leaving the conserved spin label unchanged while sending $\bk\to-\bk$.

For a two-dimensional band, $C_{2z}$ sends $\bk\to-\bk$. Spinless time reversal and $[C_{2}\Vert C_{2z}]$ therefore imply, respectively,
\begin{align}
	E_{n,s}(\bk)&=E_{n,s}(-\bk),\nonumber\\
	E_{n,s}(\bk)&=E_{n,-s}(-\bk).
	\label{eq:end_matter_spin_constraints}
\end{align}
Combining these relations gives $E_{n,s}(\bk)=E_{n,-s}(\bk)$ at every momentum. Thus every state has an opposite-spin partner at the same momentum, and $[C_{2}\Vert C_{2z}]$ is incompatible with a spin-split stripe AM whenever $[\bar{C}_{2}\Vert\mathcal{T}]$ is preserved.

{\it \color{blue}Minimal model for stripe AM.--} Motivated by the construction in Fig.~\ref{fig2}(a), we formulate a minimal model for stripe AM. Because the orbital and spin orders have the same ordering vector, the orbital order does not further enlarge the magnetic unit cell. At a phenomenological level, its redistribution of local electron density can therefore be encoded in effective hoppings, allowing us to suppress the explicit orbital degree of freedom. Retaining only sublattice and spin yields the four-band Hamiltonian
\begin{equation}
	\mathcal{H}_{\rm min}(\bk)
	=\left[\Delta_{0}^{\prime}(\bk)\sigma_{0}
	+\Delta_{x}^{\prime}(\bk)\sigma_{x}
	+\Delta_{z}^{\prime}(\bk)\sigma_z\right]s_0
	+J\sigma_zs_{z},
	\label{eq:minimal_model}
\end{equation}
where $\boldsymbol{\sigma}$ and $\boldsymbol{s}$ act in sublattice and spin space, respectively. The momentum-dependent coefficients are
\begin{align}
	\Delta_{0}^{\prime}(\bk)
	&=2\widetilde{t}_y\cos k_y+2\widetilde{t}_{x1}\cos(2k_x)\nonumber\\
	&\quad+2\widetilde{t}_{d1}\!\left[\cos(2k_x+k_y)+\cos(2k_x-k_y)\right],
	\nonumber\\
	\Delta_{x}^{\prime}(\bk)
	&=2\widetilde{t}_{x0}\cos k_x+4\widetilde{t}_{d0}\cos k_x\cos k_y,
	\nonumber\\
	\Delta_{z}^{\prime}(\bk)
	&=2\widetilde{t}_{d2}\!\left[\cos(2k_x-k_y)-\cos(2k_x+k_y)\right].
	\label{eq:minimal_model_coefficients}
\end{align}
The tildes distinguish these phenomenological hopping amplitudes from those of the model in Eq.~\ref{eq: stripeAM ele}. The terms $\Delta_{0}^{\prime}$, $\Delta_{x}^{\prime}$, and $\Delta_{z}^{\prime}$ describe sublattice-symmetric same-sublattice hopping, intersublattice hopping, and the difference between the same-sublattice hopping amplitudes on the two sublattices, respectively.

For fixed spin $s=\pm1$, the corresponding block is written as $\mathcal{H}_s(\bk)=\Delta_{0}^{\prime}\sigma_0+\Delta_{x}^{\prime}\sigma_x+(\Delta_{z}^{\prime}+sJ)\sigma_z$. In the chosen basis, $\sigma_x$ exchanges the two sublattices, while the spin part of the spin-reversing mirror exchanges $s=\pm1$. The coefficients $\Delta_{0}^{\prime}$ and $\Delta_{x}^{\prime}$ are even under $\mathcal{M}_x:(k_x,k_y)\mapsto(-k_x,k_y)$, whereas $\Delta_{z}^{\prime}$ is odd. The fixed-spin blocks therefore satisfy
\begin{align}
	\sigma_x\mathcal{H}_s(k_x,k_y)\sigma_x
	&=\mathcal{H}_{-s}(-k_x,k_y),\nonumber\\
	\sigma_x\mathcal{H}_s(\bk)\sigma_x-\mathcal{H}_{-s}(-\bk)
	&=-2\Delta_{z}^{\prime}(\bk)\sigma_z.
	\label{eq:minimal_model_symmetries}
\end{align}
The first relation explicitly establishes $[C_2\Vert\mathcal{M}_x]$. By contrast, $\Delta_{z}^{\prime}(-\bk)=\Delta_{z}^{\prime}(\bk)$, so finite $\widetilde{t}_{d2}$ breaks the candidate $[C_2\Vert C_{2z}]$ symmetry through the second relation. The Hamiltonian likewise lacks higher-fold spin-reversing rotations. For $\eta,s=\pm1$, the spectrum is
\begin{equation}
	E_{\eta,s}(\bk)=\Delta_{0}^{\prime}(\bk)
	+\eta\sqrt{\left[\Delta_{x}^{\prime}(\bk)\right]^2
	+\left[\Delta_{z}^{\prime}(\bk)+sJ\right]^2}.
	\label{eq:minimal_model_spectrum}
\end{equation}
The cross term $2sJ\Delta_{z}^{\prime}(\bk)$ produces spin splitting at generic momenta when $J\widetilde{t}_{d2}\neq0$, making explicit that both stripe exchange and sublattice-asymmetric hopping are required. The mirror relation gives $E_{\eta,s}(k_x,k_y)=E_{\eta,-s}(-k_x,k_y)$, while spinless time reversal gives $E_{\eta,s}(\bk)=E_{\eta,s}(-\bk)$. Together, they also give $E_{\eta,s}(k_x,k_y)=E_{\eta,-s}(k_x,-k_y)$. The spin splitting is therefore even under $\bk\to-\bk$ but odd under either momentum reflection. Because
\begin{equation}
	\Delta_{z}^{\prime}(\bk)
	=4\widetilde{t}_{d2}\sin(2k_x)\sin k_y,
\end{equation}
it vanishes on $k_x=0,\pm\pi/2$ and on $k_y=0$ and the equivalent Brillouin-zone boundaries. The $k_x$ nodes lie on $\mathcal{M}_x$-invariant lines and follow directly from the spin-reversing mirror. The $k_y$ nodes are likewise symmetry enforced, now by the mirror combined with spinless time reversal. The retained sine harmonic specifies only the interpolation between these nodal lines.

Equation~\eqref{eq:minimal_model_spectrum} has the same algebraic form as minimal models of Néel AMs~\cite{Roig2024AM}. Near $\Gamma$, the spin-splitting 
\begin{align}
	\Delta E_{\eta}(\bk)&\equiv E_{\eta,+}(\bk)-E_{\eta,-}(\bk)\nonumber\\
	&\simeq \frac{16\eta\widetilde{t}_{d2}Jk_xk_y}{\sqrt{(2\widetilde{t}_{x0}+4\widetilde{t}_{d0})^2+J^2}}
\end{align}
is locally $d_{xy}$-like. However, this resemblance does not furnish a global $l$-wave label, because the full stripe-anisotropic Hamiltonian lacks $[C_{2}\Vert C_{4z}]$ symmetry. A mirror, rather than a rotation, relates the opposite-spin bands and constrains the complete spin-splitting pattern.

\end{document}